\documentclass[preprint,12pt]{elsarticle}
\usepackage{amsmath}
\usepackage{amssymb}
\journal{Nuclear Physics B}

\begin{document}

\begin{frontmatter}

\title{Thermodynamics of the Topological Kondo Model} 

\author[a]{Francesco Buccheri}
\address[a]{International Institute of Physics, Universidade Federal 
do Rio Grande do Norte, 59078-400 Natal-RN, Brazil}

\author[a,b]{Hrachya Babujian}
\address[b]{Yerevan Physics Institute, Alikhanian Brothers 2,
Yerevan, 375036, Armenia}

\author[a,c]{Vladimir E. Korepin}
\address[c]{C. N. Yang Institute for Theoretical Physics, 
Stony Brook University, NY 11794, USA}

\author[a,d]{Pasquale Sodano}
\address[d]{Departemento de Fis\'ica Teorica e Experimental,
Universidade Federal do Rio Grande do Norte, 59072-970 Natal-RN, Brazil}

\author[e,f]{Andrea Trombettoni}
\address[e]{CNR-IOM DEMOCRITOS Simulation Center, Via Bonomea 265, I-34136
Trieste, Italy}
\address[f]{SISSA and INFN, Sezione di Trieste, Via Bonomea 265, I-34136 
Trieste, Italy}

\begin{abstract}
Using the thermodynamic Bethe ansatz, we investigate the topological Kondo model, which describes a set of 
one-dimensional external wires, pertinently coupled to a central region hosting a set of Majorana bound states.
After a short review of the Bethe ansatz solution, we study the system at finite temperature
and derive its free energy for arbitrary (even and odd) number of external wires.
We then analyse the ground state energy as a function of the
number of external wires and of their couplings to the Majorana bound states.
Then, we compute, both for small and large temperatures,
the entropy of the Majorana degrees of freedom localized within the central region
and connected to the external wires.
Our exact computation of the impurity entropy provides evidence of the importance of fermion
parity symmetry in the realization of the topological Kondo model.
Finally, we also obtain the low-temperature behaviour of the specific heat of the Majorana bound states,
which provides a signature of the non-Fermi-liquid nature of the strongly coupled fixed point.
\end{abstract}


\end{frontmatter}

\section{Introduction}
The possibility of faithfully simulating quantum low dimensional systems gives today the unique opportunity of testing against experiments predictions made by powerful methods -- such as bosonization \cite{giamarchi2003quantum} 
(see also Schulz et al. in \cite{morandi2000field}), conformal field theory \cite{di2009conformal}, integrable models and 
Bethe ansatz \cite{korepin1997quantum} -- developed in theoretical investigations of strongly correlated condensed matter and spin systems.
As a result, one may now confidently apply the above methods to newly engineered quantum systems relevant for the fabrication of quantum devices as well as for the realization, in a easily controllable setting, of new phases of matter such as the non-Fermi liquid phases realized in Dirac materials \cite{Balatsky} and in overscreened multi-channel Kondo models \cite{Schlottmann1993}.  Within the new models made accessible to theoretical investigations, great opportunities are offered by the analysis of pertinent junctions of one dimensional systems such as ladders and star junctions \cite{HoudePhysRevLett.85.5543,Schumm05,SchummKruger,Gattobigio,Atala14,argawal2014}.

Networks of one dimensional models attracted much attention over the past few years. In pioneering works \cite{ChamonPhysRevLett.91.206403,Chamon} a star junction of three quantum wires enclosing a magnetic flux was studied: modelling the wires as Tomonaga-Luttinger liquids (TLL), the authors of Ref. \cite{ChamonPhysRevLett.91.206403,Chamon} were able to show the existence of an attractive finite coupling fixed point, characteristic of the geometry of the circuit. Later, a repulsive finite coupling fixed point was found in a T-junction of one dimensional Bose liquids \cite{TokunoPhysRevLett.100.140402}. Crossed TLL were also the subject of many investigations both analytical \cite{Kazymyrenko} and numerical \cite{Guo}: these analyses pointed out that, in crossed TLLs, the junction induces behaviours similar to those arising from quantum Kondo impurities in condensed matter physics \cite{AffleckLesHouches}. For what concerns the analysis of junctions of spin chains, in ref. \cite{Reyes} it was argued that 
novel 
critical behaviours emerge when crossing at a point two spin ${1}/{2}$ Heisenberg models since, as a result of the crossing, some operators turn from irrelevant to marginal, leading to correlation functions exhibiting power law decays with non-universal exponents. Star junctions of Josephson junction arrays were investigated in \cite{Giuliano} with the result that a finite coupling fixed point was also emerging in these superconducting systems. With Majorana fermions \cite{WilczeckMajoranaReturns} star junctions of quantum wires become very attractive since these geometries facilitate their braiding \cite{Alicea} allowing, at least in principle, for the engineering of quantum circuits relevant for the implementation of quantum protocols \cite{NielsenChuang,NayakTQC}. 

The very close relation between the phase diagram emerging from the investigation of networks of quantum one-dimensional systems (quantum spin chains and quantum wires, essentially) and the one typical of multichannel Kondo models was established only very recently. In the two papers \cite{Tsvelik:2014Ising,Crampe2013} it was shown that a star junction of three critical Ising models and a star junction of three XX models may be made equivalent to the two channel and the four channel over-screened Kondo model, respectively. To achieve their exact mapping, these authors used a generalization of the Jordan-Wigner transformation needed to satisfy the anticommutation relations between fermions located on different legs of the junction. For this purpose, one modifies the usual Jordan-Wigner transformation  by the addition of an auxiliary space made - for a star junction of three spin chains - by three Klein factors, i.e. three real anti-commuting fields, lying at the inner boundary of each chain.
As a result, in these realizations of the multichannel Kondo model, the central spin, with which the Jordan-Wigner fermions interact, is realized as a
non-local
combination of three Klein factors.
The Kondo effect occurring when bulk fermions scatter on a composite ``spin''
non-locally 
encoded by any number of Klein factors located at different space points provides a realization of the topological Kondo effect.
Since one expects that an extended
spin is less sensitive to noise and decoherence, it is generally believed that this realization of Kondo models could 
``naturally`` be much more robust than the one realized by other means (for example with quantum dots \cite{PotokNature}).

\begin{figure}[h!]
 \centering
 \includegraphics[width=0.6\textwidth]{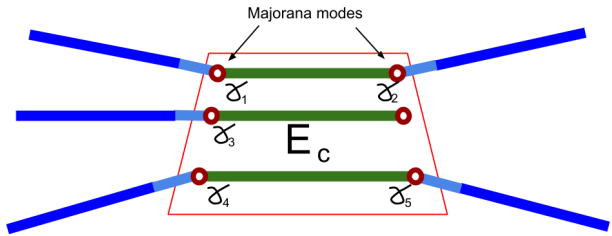}
 \caption{Schematic representation of the device realizing the topological Kondo effect: Majorana edge modes of quantum wires are hosted on an s-wave superconductor, capacitively coupled with the ground. 
 These modes are proximity-coupled with the external leads.}
\label{fig:MajoranaBox}
\end{figure}

In an effort to look for new experimental realization of a 
two-channel Kondo Model the authors of \cite{BeriCooper2012,AltlandEgger} 
showed that, in pertinent circuit regimes, networks of quantum wires supporting edge Majorana modes \cite{ZazunovEvenOdd} provide a first experimentally attainable realization of the so-called Topological Kondo effect. Subsequently, the spin dynamics \cite{Altland2013}, as well as the exact solution for finite number of electrons \cite{Altland2014}, was investigated.
In realistic cases the topological Kondo effect may be realized in networks of quantum wires supporting edge Majorana modes.
For instance, it takes place when a mesoscopic superconducting island, capacitively coupled with the ground,
hosts a set of Majorana bound states, which may be the Majorana edge modes (MEMs) of a set of spin-orbit coupled nanowires (green in figure \ref{fig:MajoranaBox} in colour version)
laying on this superconductor.
Note that, in this realization, the MEMs always appear in pairs, localized at the opposite ends of the nanowire.
As a result, the central region of a circuit realizing the topological Kondo effect contains always an even number of MEMs.
Nevertheless, it is always possible to couple only an odd number of MEMs to external wires,
thus leaving an odd number of Majorana modes decoupled from the rest of the system.

Coulomb interactions among the electrons within the external leads are not essential for the topological Kondo effect
and will not be accounted for in the following. However, for the host superconducting region,
they play an essential role, since they determine its charging energy $E_c$ \cite{BeriCooper2012,Altland2013}.
We focus on the temperature regime $T\ll E_c$, which implies that, for all real processes, the number
of electrons on the island is conserved.
We emphasize that, at least at low temperature, the MEMs must be the only degrees of freedom
in the central region which are involved in the dynamics.
Under these conditions, the effective low-energy Hamiltonian describing the TKM is \cite{BeriCooper2012,AltlandEgger,Beri2013,Zazunov14}
\begin{eqnarray}\label{eq:TopologicalKondoHamiltonian}
 H&=&-i\frac{\hbar v_{F}}{2\pi}\sum_{\alpha=1}^{M}\intop dx \psi_{\alpha}^{\dagger}(x)\partial_{x}\psi_{\alpha}(x)
 \nonumber\\
 && + \sum_{\alpha\ne\beta}\lambda_{\alpha\beta}\gamma_{\alpha}\gamma_{\beta}\psi_{\alpha}^{\dagger}(0)\psi_{\beta}(0)
 +i\sum_{\alpha\ne\beta}h_{\alpha\beta}\gamma_{\alpha}\gamma_{\beta}
 \;
\end{eqnarray}
and, manifestly, preserves fermion parity.
Here $\psi_{\alpha}(x)$ are the (complex) Fermi fields describing electrons in the wires $\alpha=1,\ldots,M$
and satisfying canonical anticommutation relations
\begin{equation}
 \left\lbrace\psi_{\alpha}(x),\psi_{\beta}(y)\right\rbrace = 0 \qquad\qquad \left\lbrace\psi_{\alpha}(x),\psi_{\beta}^\dagger (y)\right\rbrace = \delta_{\alpha,\beta}\delta(x-y)
 \;,
\end{equation}
 while the $\gamma_{\alpha}=\gamma_{\alpha}^\dagger$ are Majorana fields constrained in a box connected with the wires and satisfying the Poisson algebra
 \begin{equation}\label{eq:PoissonAlgebra}
  \left\lbrace\gamma_{\alpha},\gamma_{\beta}\right\rbrace = 2 \delta_{\alpha,\beta}
  \;.
 \end{equation}
The symmetric matrix $\lambda_{\alpha,\beta}>0$ is the analogue of the coupling with the magnetic impurity in the usual Kondo problem.
The couplings $h_{\alpha,\beta}=h_{\beta,\alpha}$ represent a direct interaction between only a pair of Majorana fermions
and can be made exponentially small by considering the MEMs sufficiently separated in space \cite{SemenoffSodano}.
This model coincides with the one studied in \cite{Altland2014}. A related, yet different model, with real spinless
fermions in the bulk, has been analysed in \cite{TsvelikIsing}.

If $M$ of the MEMs are coupled to the external wires, this set of MEMs 
may be regarded as a "Majorana spin": in contrast to a conventional spin the ''Majorana spin'' is
non-locally encoded in the spatially separated MEMs and governed by the orthogonal symmetry group $SO(M)$.
It has been found \cite{BeriCooper2012,AltlandEgger,Beri2013,Zazunov14} that the topological Kondo model (TKM)
provides a natural realization of the quantum critical point described by a Wess-Zumino-Witten-Novikov boundary 
conformal field theory \cite{Affleck:1990zd,Affleck:1990by,Affleck1991641,di2009conformal} and support non-Fermi-liquid ground states.

In this paper, using the thermodynamic Bethe ansatz (TBA),
we investigate the thermodynamics of the TKM with an arbitrary
number of wires $M$.
For this purpose, we study a system with a thermodynamically large number of electrons
in contact with an heat bath at temperature $T$ and at fixed density and perform the thermodynamic
limit as in (\ref{thermodynamiclimit}).
Our analysis allows to derive the free energy of the TKM for arbitrary number of external wires,
to compute the ground state energy as a function of the number of wires and of their couplings to the 
Majorana modes, as well as the low temperature behaviour of the specific heat of the central region.

An outline of the structure of the paper and of our results 
is as follows: In section \ref{sec:The-Bethe-ansatz} 
we review the exact solution of the TKM, as provided in \cite{Altland2014}.

In section \ref{sec:TBA} we derive 
the exact free energy of the TKM made by $M$ wires coupled with $M$
MEMs in the central region.
Here, we write the TBA equations for both even and odd $M$.
In section \ref{sec:GSenergy} we obtain, in a closed form, the exact ground state energy shift
due to the presence of the Majorana modes in the central region
\begin{equation}\label{eq:GroundStateNRG}
E_{J}^{(0)}=E_{J}^{(0)}(\lambda,M)=i\log\frac{i\Gamma\left(\frac{M+2}{4(M-2)}+\frac{i}{(M-2)\lambda}\right)\Gamma\left(\frac{3M-2}{4(M-2)}-\frac{i}{(M-2)\lambda}\right)}{\Gamma\left(\frac{M+2}{4(M-2)}-\frac{i}{(M-2)\lambda}\right)\Gamma\left(\frac{3M-2}{4(M-2)}+\frac{i}{(M-2)\lambda}\right)}
\end{equation}
as a function of the effective tunnelling strength $\lambda$ between the wires.

In section \ref{sec:Entropy}, we compute the entropy associated with the Majorana 
degrees of freedom localized in the central region both for $T\to0$ and $T\to\infty$.
This will provide us with the exact expression of the impurity entropy of the TKM, in these two relevant limits.
As expected, for $T\to\infty$, the computation of the entropies amounts to determine
the dimension of the Hilbert space $\mathcal{H}_J(M)$ associated with the
MEMs localized in the central region and coupled with the external wires of the circuit. We find:
\begin{equation}\label{eq:dimensionImpurityHilbert}
  \dim\mathcal{H}_J(M)=2^{\left\lfloor\frac{M-1}{2}\right\rfloor}
  \;,
\end{equation}
which summarizes equations (\ref{eq:DK-dimensionImpurityHilbert}) and (\ref{eq:BK-dimensionImpurityHilbert}) 
for even and for odd values of $M$, respectively.
Equation (\ref{eq:dimensionImpurityHilbert}) shows that the dimension of the impurity Hilbert space
of $M$ wires for $M$ odd equals the dimension of the Hilbert space of a system with $M+1$ (even) wires.
As a result, for $T\to\infty$ and for any given $M$ it is always possible to preserve fermion parity symmetry
since \cite{SemenoffSodano} with an even number of Majorana modes one can always make the lowest-energy 
eigenstate of (\ref{eq:TopologicalKondoHamiltonian}) an eigenstate of the fermion number parity.
As a result, the dimension $\mathcal{H}_J$ corresponds to the dimension of $M$ Majorana modes projected to a 
definite fermion number parity sector.
Remarkably, this result, stemming only from TBA, is consistent with the approach used in \cite{Altland2014},
where the projection to a definite fermion parity sector was needed to ensure the equivalence of the TKM for specific
values of $M$ to multi-channel Kondo and Coqblin-Schrieffer models.

For the analysis of the $T\to0$ limit of the impurity entropy, one should recall \cite{AffleckGroundState} that,
in one-dimensional problems (or effectively one-dimensional, as in the present case)
with boundaries or impurities, the general form of the logarithm of the partition function of a critical
system of size $\mathcal{L}$ at temperature $T$ behaves like
\begin{equation}\label{eq:criticallogZ}
 \log Z \sim \frac{\pi \mathcal{L} T c}{6} + S^{(0)}\,;
\end{equation}
In eq. (\ref{eq:criticallogZ}), $\mathcal{L}\ll 1/T$, $c$ is the central charge 
and characterizes the conformal field theory describing the low-temperature critical
point and $S^{(0)}=\log g$, where $g$ is the ''ground state degeneracy``, is a lengt-independent
term which may appear without breaking conformal invariance. Remarkably, $g$ is a universal quantity
which can be a non-integer number for $\mathcal{L}\to\infty$.
The computation of $S^{(0)}$ is given in section \ref{sec:Entropy} using the TBA in the limit $T\to 0$ and yields:
\begin{equation}\label{eq:residualEntropy}
   S^{(0)}_J=\log \sqrt{\frac{M}{2}}\qquad(\mbox{even }M)\; ,\qquad\qquad S^{(0)}_J=\log \sqrt{M}\qquad(\mbox{odd }M)
\end{equation}
This expression summarizes equations (\ref{eq:DK-residualEntropy}) and (\ref{eq:BK-residualEntropy})
and agrees with the conformal perturbation theory \cite{AffleckGroundState} predictions given, for any $M$, in \cite{Altland2014}.

Finally, we show in section \ref{sec:SpecificHeat} that the system exhibits non-Fermi liquid
behaviour at low temperatures by computing the temperature dependence of the specific heat of
the Majorana modes in the central region.
We find that the power expansion exhibits a term proportional to $T^{2\frac{M-2}{M}}$.
We argue that the power $2(M-2)/M$ originates only from the symmetry of the strong-coupling fixed point.

\section{The Bethe ansatz solution}\label{sec:The-Bethe-ansatz}

The diagonalization of (\ref{eq:TopologicalKondoHamiltonian}) has been
performed in \cite{Altland2014} and it is based on the overall $SO(M)$
symmetry of the problem and on the results of \cite{Ogievetsky1987}
(see also \cite{DeVegaO2N}).
The exact solution has been constructed using the general symmetry $SO(M)$
and the equivalence of the cases with $M=3, 4, 6$ wires
to some impurity problems whose diagonalization had been performed earlier
\cite{Tsvelick85,Andrei1984,Schlottmann1993,Jerez1998}.

The Bethe Ansatz provides the exact quantized momenta $\left\lbrace k_n \right\rbrace_{n=1,\ldots,N}$
of the bulk electrons as a function of the solution of a set of nested Bethe ansatz equations. 
The energy is read from the eigenvalues of the transfer matrix \cite{Ogievetsky1987} as: 
\begin{equation}
E
=\frac{v_F\hbar}{2} \sum_{j=1}^N k_j
=\frac{v_{F}\hbar}{2i\mathcal{L}}\sum_{j=1}^{r_{1}}\log e_{2}(x_{j}^{(1)})
=\frac{2E_{F}}{i\pi}\frac{M}{N}\sum_{j=1}^{r_{1}}\log e_{2}(x_{j}^{(1)})
\label{eq:DK-Energy}
\;,
\end{equation}
where
\begin{equation}
p_{F}=
mv_{F}=\frac{\hbar\pi N}{2\mathcal{L}M}
\;,
\qquad
E_{F}=\frac{p_{F}^{2}}{2m}\propto\frac{1}{M^{2}}
\;,
\qquad
e_{n}(x)=\frac{x-in/2}{x+in/2}\label{eq:endef}
\end{equation}
and $\mathcal{L}$ is the length of each wire.
The set of rapidities $\left\lbrace x^{(1)}_j \right\rbrace$ satisfies,
for even number of branches $M=2K$, 
the system of coupled algebraic equations (Bethe equations):
{\small
\begin{eqnarray}
\left[e_{2}\left(x_{j}^{(1)}\right)\right]^{N} & = & 
\frac{\prod_{k=1}^{r_{1}}e_{2}\left(x_{j}^{(1)}-x_{k}^{(1)}\right)}
{\prod_{k=1}^{r_{2}}e_{1}\left(x_{j}^{(1)}-x_{k}^{(2)}\right)}\nonumber \\
1 & = & \frac{\prod_{k=1}^{r_{l}}e_{2}\left(x_{j}^{(l)}-x_{k}^{(l)}\right)}
{\prod_{k=1}^{r_{l-1}}e_{1}\left(x_{j}^{(l)}-x_{k}^{(l-1)}\right)
\prod_{k=1}^{r_{l+1}}e_{1}\left(x_{j}^{(l)}-x_{k}^{(l+1)}\right)}
\quad 1<l<K-2\nonumber \\
1 & = & \frac{\prod_{k=1}^{r_{K-2}}e_{2}\left(x_{j}^{(K-2)}-x_{k}^{(K-2)}\right)
\prod_{k=1}^{r_{K-3}}e_{-1}\left(x_{j}^{(K-2)}-x_{k}^{(K-3)}\right)}
{\prod_{k=1}^{r_{K}}e_{1}\left(x_{j}^{(K-2)}-x_{k}^{(K)}\right)
\prod_{k=1}^{r_{K-1}}e_{1}\left(x_{j}^{(K-2)}-x_{k}^{(K-1)}\right)
}\nonumber \\
e_{1}\left(x_{j}^{(K-1)}-\frac{1}{\lambda}\right) & = & \frac{\prod_{k=1}^{r_{K-1}}e_{2}\left(x_{j}^{(K-1)}-x_{k}^{(K-1)}\right)}{\prod_{k=1}^{r_{K-2}}e_{1}\left(x_{j}^{(K-1)}-x_{k}^{(K-2)}\right)}\nonumber \\
1 & = & \frac{\prod_{k=1}^{r_{K}}e_{2}\left(x_{j}^{(K)}-x_{k}^{(K)}\right)}{\prod_{k=1}^{r_{K-1}}e_{1}\left(x_{j}^{(K)}-x_{k}^{(K-1)}\right)}
\label{eq:DK-BE}
\;.
\end{eqnarray}
}
The set of integers $r_{1},\ldots,r_{K}$, denoting the number
of roots in each level of the nested Bethe ansatz, defines a sector
corresponding to given eigenvalues of the Cartan operators of the lie 
algebra $D_K$, generating the group $SO(M)$ with even $M=2K$
\cite{Ogievetsky1987}.

For an odd number of branches $M=2K+1$, the energy is given again 
by (\ref{eq:DK-Energy}), but the rapidities $\left\{ x_{j}^{(1)}\right\} _{j=1,\ldots,r_{1}}$
solve now the system \cite{Ogievetsky1987}: 
{\small
\begin{eqnarray}\label{eq:BA-BK}
\left[e_{2}\left(x_{j}^{(1)}\right)\right]^{N}\prod_{k=1}^{r_{2}}e_{1}\left(x_{j}^{(1)}-x_{k}^{(2)}\right) & = & \prod_{k=1}^{r_{1}}e_{2}\left(x_{j}^{(1)}-x_{k}^{(1)}\right)\nonumber \\
\prod_{k=1}^{r_{l-1}}e_{1}\left(x_{j}^{(l)}-x_{k}^{(l-1)}\right)\prod_{k=1}^{r_{l+1}}e_{1}\left(x_{j}^{(l)}-x_{k}^{(l+1)}\right) & = & \prod_{k=1}^{r_{l}}e_{2}\left(x_{j}^{(l)}-x_{k}^{(l)}\right)
\qquad1<l<K \nonumber\\
e_{1/2}\left(x_{j}^{(K)}-\frac{1}{\lambda}\right)\prod_{k=1}^{r_{K-1}}e_{1}\left(x_{j}^{(K)}-x_{k}^{(K-1)}\right) & = & \prod_{k=1}^{r_{K}}e_{1}\left(x_{j}^{(K)}-x_{k}^{(K)}\right) \quad,
\end{eqnarray}
}
associated this time with the non simply-laced $B_{K}$ algebra,
generating the group $SO(M)$ with odd $M=2K+1$.
We refer to \cite{Tsvelik:2014Ising} for the Bethe equations of this model. 

In the following section, we will use the notation and the results shown
here to derive the thermodynamics of the TKM.

\section{The thermodynamic Bethe ansatz}\label{sec:TBA}

In this section, we study the temperature dependence of the
 free energy of the system by means of the thermodynamic Bethe
 ansatz, which will allow for considerably simpler exact 
 expressions for the thermodynamic quantities when the system
 size is large.
 The thermodynamic limit is defined as:
\begin{equation}
N\to\infty,\qquad\mathcal{L}\to\infty,\qquad N/\mathcal{L}=const \;.
\label{thermodynamiclimit}
\end{equation}
In this limit the roots
 of the system (\ref{eq:DK-BE}), are known to group into clusters
 of roots with the same real part and equally spaced imaginary parts,
 called ``string'' solutions of the Bethe equations.

\subsection{Even number of branches: M=2K}\label{sub:Even-number-of}

In this case, the roots at each level group as follows:
\begin{equation}\label{eq:WideString}
\left\{ x_{n;\alpha}^{(j)}+\frac{i}{2}\left(n-2l+1\right):x_{n;\alpha}^{(j)}\in\mathbb{R},\; l=1,\ldots,n\quad j=1,\ldots,K\right\} 
\;,
\end{equation}
in which the real number $x_{n;\alpha}^{(j)}$ is the string ``center'',
i.e., the common real part of the solution, the index $j$ denotes
the level, the subscripts $n$ and $\alpha$ denote the string index
and the root index within the bound state.

The number of such clusters goes to infinity, while the spacing of
the string centers goes to zero in the limit of infinite number of
sites: therefore, a more convenient description is given in terms
of densities of solutions. In particular, we will denote by $\rho_{n}^{(j)}$
 the density of solutions of length $n$ at the $j$-th level of
 the nesting and by $\tilde{\rho}_{n}^{(j)}$ the density of unoccupied levels
 (holes) for the same type of configurations.
 
We first substitute the form (\ref{eq:WideString}) for the solutions of the
Bethe equations into (\ref{eq:DK-BE}), then we group the terms of
the products in the RHS according to their real part (string center).
Then, we multiply all the equations corresponding to the different roots
in the same string among them and we consider $-i$ times the logarithm of
the resulting equation.
The kernel associated with the ``scattering'' of an $n$-string
and an $m$-string is (hat denotes Fourier transform):
\begin{equation}
\hat{A}_{n,m}(\omega)=\coth\frac{|\omega|}{2}\left(e^{-|n-m|\frac{|\omega|}{2}}-e^{-(n+m)\frac{|\omega|}{2}}\right)
\label{eq:AnMatrix}
\;.
\end{equation}
On the other hand, the logarithmic derivative of the function $e_{1}$ produces
the kernel $s*A_{n,m}$, where the symbol $*$ denotes convolution
\begin{equation}
 \left(f*g\right)(x)=\intop_{-\infty}^{\infty} dy f(x-y)g(y)
\end{equation}
and
\begin{equation}
\hat{s}(\omega)=\frac{1}{2\cosh\frac{\omega}{2}}  \; .\label{eq:s}
\end{equation}
The resulting density equations are: 
\begin{eqnarray}
2\pi\tilde{\rho}_{n}^{(j)}(x)&=&\delta^{j,1}\left((1-\delta_{n,1})a_{n-1}+a_{n+1}\right)(x)+\delta^{j,K-1}\frac{1}{N}a_{n}\left(x-\frac{1}{\lambda}\right)   \nonumber\\
&&-\left(M^{j,l}*A_{nm}*\rho_{m}^{(l)}\right)(x)
\label{eq:DK-densityequation}
\;,
\end{eqnarray}
in which there appear the function $a_{n}(\omega)=e^{-n\frac{|\omega|}{2}}$
and the $K\times K$ matrix
\begin{equation}
M=\left(\begin{array}{cccccc}
1 & -s\\
-s & 1 & \ddots\\
 & \ddots & \ddots & -s\\
 &  & -s & 1 & -s & -s\\
 &  &  & -s & 1\\
 &  &  & -s &  & 1
\end{array}\right)\label{eq:DK-C}
\end{equation}
encodes the structure of the $D_{K}$ algebra.

It is possible to write the free energy at temperature
$\tau$ in the usual way as 
\begin{equation}\label{eq:freeNrgCanonical}
 F=E-\tau S
 \;,
\end{equation}
where $S$ is the entropy of the whole system and is computed counting
the number of states associated to a given root density as (sum over
repeated indexes is implied):
\begin{eqnarray}
S&=&\sum_{j=1}^K\sum_{m=1}^\infty
\intop_{\mathbb{R}}dx \Big[\left(\rho_{m}^{(j)}+\tilde{\rho}_{m}^{(j)}\right)\log\left(\rho_{m}^{(j)}+\tilde{\rho}_{m}^{(j)}\right)(x)
\nonumber\\
&&\qquad\qquad\qquad-\rho_{m}^{(j)}(x)\log\rho_{m}^{(j)}(x)-\tilde{\rho}_{m}^{(j)}(x)\log\tilde{\rho}_{m}^{(j)}(x) \Big]
\;.
\end{eqnarray}
The partition function will be dominated by the root configurations
corresponding to the densities that make the free energy
stationary in the thermodynamic limit.
These can be selected by imposing the variation of (\ref{eq:freeNrgCanonical})
with respect to $\rho_{n}^{(j)}$ to be vanishing, also taking into account 
the constraint (\ref{eq:DK-densityequation}).
Defining
\begin{equation}\label{eq:PhiDef}
e^{\phi_{n}^{(j)}(x)} \equiv \frac{\rho_{n}^{(j)}(x)}{\tilde{\rho}_{n}^{(j)}(x)}
\;,
\end{equation}
such configurations are characterized by the saddle point equations:
\begin{eqnarray}
0 & = & \delta^{j,1}\bar{C}_{nm}*E_{m}^{(1)}-T\,\bar{C}_{nm}*L_{-,m}^{(j)}+T\, M^{jl}*L_{+,n}^{(l)}\label{eq:DK-SaddlePoint}
\end{eqnarray}
with $\hat{E}_{n}^{(1)}(\omega)=\frac{a_{n-1}(\omega)+a_{n+1}(\omega)}{i\omega}-2\pi\delta(\omega)$
for $n>1$, $\hat{E}_{1}^{(1)}(\omega)=\frac{a_{2}(\omega)}{i\omega}-\pi\delta(\omega)$
and $L_{\pm,m}^{(j)}(x)=\log\left(1+e^{\pm\phi_{m}^{(j)}(x)}\right)$.
We have chosen to measure all the energies rescaled by $\frac{2E_{F}}{\pi}$
and defined the dimensionless temperature $T=\frac{\pi k_{B}\tau}{2E_{F}}$
($k_{B}$ is the Boltzmann constant). The matrix:
\begin{equation}
\left[\bar{C}\right]_{mn}=\left[A^{-1}\right]_{mn}=\delta_{mn}-\hat{s}(\omega)\left(\delta_{m+1,n}+\delta_{m-1,n}\right)\label{eq:Cbar}
\end{equation}
is the inverse of (\ref{eq:AnMatrix}). The saddle point equations
can be put in a more explicit form as (assume $K\ge3$):
\begin{eqnarray}\label{eq:DK-phi-numerics}
\phi_{n}^{(j)}(x) & = & \frac{1}{T}\arctan\left(e^{\pi x}\right)\delta^{j,1}\delta_{n,2}
\nonumber \\
&&
-s*\left[L_{-,n-1}^{(j)}+L_{-,n+1}^{(j)}-L_{+,n}^{(j-1)}-L_{+,n}^{(j+1)}\right] (x)
\qquad\; j<K-2\nonumber \\
\phi_{n}^{(K-2)}(x) & = & -s*\left[L_{-,n-1}^{(K-2)}+L_{-,n+1}^{(K-2)}-L_{+,n}^{(K-3)}-L_{+,n}^{(K-1)}-L_{+,n}^{(K)}\right](x) \\
\phi_{n}^{(j)}(x) & = & -s*\left[L_{-,n-1}^{(j)}+L_{-,n+1}^{(j)}-L_{+,n}^{(K-2)}\right](x)
 \qquad\qquad  j=K-1,K
\;.
\nonumber 
\end{eqnarray}
The case $K=2$ ($M=4$) is instead written as:
\begin{eqnarray*}
\phi_{n}^{(1)}(x) & = & \frac{1}{T}\arctan\left(e^{\pi x}\right)\delta_{n,2}-s*\left[L_{-,n-1}^{(1)}+L_{-,n+1}^{(1)}-L_{+,n}^{(2)}\right](x)\\
\phi_{n}^{(2)}(x) & = & -s*\left[L_{-,n-1}^{(2)}+L_{-,n+1}^{(2)}-L_{+,n}^{(1)}\right](x)
\;.
\end{eqnarray*}
With the aid of these functions, it is now possible to write the free
energy in a more compact way. It differs from the free energy of $M$
uncoupled wires by terms which are $O\left(\frac{1}{N}\right)$,
which can therefore be associated with the Majorana degrees of freedom 
in the central region and dubbed "junction" free energy.
We write it as:
\begin{equation}\label{eq:DK-FreeNRG-long}
F_{J}=-T\sum_{m=1}^{\infty}\intop_{\mathbb{R}}dxa_{m}\left(x-\frac{1}{\lambda}\right)L_{+,m}^{(K-1)}(x)
\;.
\end{equation}
The latter expression contains an infinite sum over the string lengths
$m$ and is therefore not very convenient for numerical evaluation.
Hence, we multiply the system (\ref{eq:DK-phi-numerics}) by the matrix
$\tilde{A}$, i.e., the inverse of the matrix $M$ defined in
(\ref{eq:DK-C}) and extract an equality for $L_{+,m}^{(j)}(x)$.
In the case $K=2$ the matrix $\tilde{A}$ reads:
\begin{eqnarray}\label{eq:DK-AtildeK2}
 \hat{\tilde{A}}^{1,1}(\omega)=\hat{\tilde{A}}^{2,2}(\omega)&=&  2\frac{1+\cosh\omega}{1+2\cosh\omega} \nonumber\\
  \hat{\tilde{A}}^{1,2}(\omega)=\hat{\tilde{A}}^{2,1}(\omega)&=& \frac{\sinh\omega}{\sinh\frac{3\omega}{2}}
  \;.
\end{eqnarray}
For higher $K$, one has the symmetric matrix:
\begin{eqnarray}
\mbox{for }j,l<s-1: \quad 
\hat{\tilde{A}}^{j,l} & = & \frac{2\cosh\frac{\omega}{2}\cosh\frac{(K-1-\max\left(j,l\right))\omega}{2}}{\cosh\frac{(K-1)\omega}{2}}\frac{\sinh\frac{\min(j,l)\omega}{2}}{\sinh\frac{\omega}{2}}\qquad \nonumber \\
\hat{\tilde{A}}^{j,K-1}=\hat{\tilde{A}}^{j,K} & = & \frac{\cosh\frac{\omega}{2}}{\cosh\frac{(K-1)\omega}{2}}\frac{\sinh\frac{j\omega}{2}}{\sinh\frac{\omega}{2}}\qquad j<K-1\nonumber \\
\hat{\tilde{A}}^{K-1,K} & = & \frac{1}{2\cosh\left(\frac{(K-1)\omega}{2}\right)}\frac{\sinh\frac{\left(K-2\right)\omega}{2}}{\sinh\frac{\omega}{2}}\nonumber \\
\hat{\tilde{A}}^{K,K}=\hat{\tilde{A}}^{K-1,K-1} & = & 1+\frac{1}{2\cosh\left(\frac{(K-1)\omega}{2}\right)}\frac{\sinh\frac{\left(K-2\right)\omega}{2}}{\sinh\frac{\omega}{2}}\label{eq:DK-Atilde}
\;.
\end{eqnarray}
This matrix has already appeared in \cite{Babichenko:2003rf} for
the TBA associated to an integrable quantum field theory with
$\frac{\mathcal{G}_{l}\times\mathcal{G}_{m}}{\mathcal{G}_{l+m}}$
symmetry, apart from an overall normalization. 
Substituting into (\ref{eq:DK-FreeNRG-long}),
we obtain the final result for the junction free energy in
a more compact form: 
\begin{eqnarray}
F_{J}(T)  
&= & E_{J}^{(0)}-T\sum_{l=1}^K\intop_{\mathbb{R}}\frac{d\omega}{2\pi}e^{i\omega/\lambda}\hat{s}\left(\omega\right)\hat{\tilde{A}}^{K-1,l}\left(\omega\right)\hat{L}_{-,1}^{(l)}\left(\omega\right)
\label{eq:DK-JunctionFreeEnergy}
\;,
\end{eqnarray}
where 
\begin{equation}\label{eq:DK-GSnrg}
E_{J}^{(0)}=\intop_{\mathbb{R}}d\omega\frac{e^{-|\omega|}\sin\frac{\omega}{\lambda}}{2\omega\cosh\frac{(M-2)\omega}{4}}-\frac{\pi}{2}
\end{equation}
is the Majorana contribution to the ground state energy (see also
\cite{Andrei83,Jerez1998}). The case of $K=3$ ($M=6$) is the one considered in \cite{Altland2014}.
The formula (\ref{eq:DK-JunctionFreeEnergy}) is valid for the topological
Kondo with any even number of branches.

\subsection{Odd number of branches: M=2K+1}\label{sub:Odd-number-of}

We present now the thermodynamic Bethe ansatz for a system with
an odd number of concurring branches. 
The case $K=1$, i.e., three wires connected with three MEMs in the central region,
is equivalent to the four-channel Kondo model
\cite{Crampe2013}. Its Bethe equations and 
thermodynamics have been studied in \cite{Andrei1984}.
The case $K=2$ is somewhat special, as there are only two levels
present, and can be read in \cite{Altland2014}. 
The right-hand side of the last of the equations (\ref{eq:BA-BK})
implies that, in the thermodynamic limit, the $K$-th level roots
to group into ``close'' strings such as:
\begin{equation}
\left\{ x_{n;\alpha}^{(K)}+\frac{i}{4}\left(n-2j+1\right)\,:\quad x_{n;\alpha}^{(K)}\in\mathbb{R},\; j=1,\ldots,n\right\} 
\label{eq:CloseStrings}
\;.
\end{equation}
For all the other levels, the strings are of the ``wide'' type (\ref{eq:WideString}).

Once again, we substitute the form of the solutions (\ref{eq:WideString}) and (\ref{eq:CloseStrings})
into (\ref{eq:BA-BK}), group the terms in the RHS according to their real part
and multiply among them all the equations which show a root of the same string 
solution in the RHS.
We then take the logarithm and multiply by $-i$ and, when taking the 
thermodynamic limit, rewrite the resulting equations in terms of string densities as
\begin{eqnarray}\label{eq:BK-densityequations}
2\pi\tilde{\rho}_{m}^{(j)}(x) & = & \delta^{j,1}\left((1-\delta_{m,1})a_{m-1}+a_{m+1}\right)\left(x\right)+\frac{\delta^{j,K}}{N}a_{m/2}\left(x-\frac{1}{\lambda}\right)
\nonumber\\
&&\qquad -\left(M^{j,l}*A_{m,n}^{(j,l)}*\rho_{n}^{(l)}\right)\left(x\right)
\;,
\end{eqnarray}
with the matrices (in Fourier transform)
\begin{equation}
\hat{A}_{m,n}^{j,l}(\omega)=\hat{A}_{\frac{m\,t_{j,l}}{t_{j}},\frac{n\, t_{j,l}}{t_{l}}}\left(\frac{\omega}{t_{j,l}}\right)\label{eq:BK-A}
\end{equation}
and
\begin{equation}
\hat{M}^{j,l}\left(\omega\right)=\delta^{j,l}-\hat{s}_{1/t_{j,l}}\left(\delta^{j,l-1}+\delta^{j,l+1}\right)
\;,
\end{equation}
in which $t_{j}=1+\delta_{j,K}$ represents the inverse length of
the $j$-th root of the $B_{K}$ algebra and is equal to one for all
the long roots and to $2$ for the short root associated to the $K$-th
node, while $t_{j,l}=\max\left(t_{j},t_{l}\right)$ and 
$\hat{s}_x=\hat{s}_x(\omega)=\hat{s}\left(x\omega\right)$.

The thermodynamics is not very different from the one for even number of wires, overall,
and pass through the computation of the variation of the free energy
(\ref{eq:freeNrgCanonical}) with respect to the string densities $\rho_m^{(j)}$
with the constraint (\ref{eq:BK-densityequations}).
Here, however, the root length and the string length are not independents, hence
we have a more complicated matrix in the equations for the sting densities,
which implies that one will have more than one frequency appearing in 
the Fourier transform of (\ref{eq:BK-densityequations}).
For this reason, we shall be rather pedantic in the subsequent exposition,
even if similar results are already available in literature
\cite{Tsvelick85,Schlottmann1991,Babichenko:2003rf,Andrei1984,Tsvelik:2014Ising}.

From the minimization of the free energy of the system at the effective
temperature $T$ (defined in section \ref{sub:Even-number-of}) we obtain
a saddle point condition for the function (\ref{eq:PhiDef}):
\begin{equation}
\delta^{j,1}E_{m}^{(1)}-TL_{-,m}^{(j)}+TM^{j,l}*A_{m,n}^{(j,l)}*L_{+,n}^{(l)}=0
\label{eq:BK-SaddlePoint}
\;.
\end{equation}
The resulting free energy is (the units
and the dimensionless temperature have been
defined in section \ref{sub:Even-number-of})
\begin{equation}
F=-T\int dx\left[\delta^{j,1}\left(s*A_{m,2}\right)\left(x\right)+\delta^{j,K}\frac{1}{N}a_{m}\left(2\left(x-\frac{1}{\lambda}\right)\right)\right]L_{+,m}^{(j)}(x)
\label{eq:BK-freeNrg-long}
\;,
\end{equation}
where the second term, subleading in the number of particles, represents
the free energy of the MEMs located at the junction. This is a rather inconvenient expression,
for we have an infinite sum over all the string lengths. In order
to simplify it, we define the matrices:
\begin{equation}
\hat{Q}_{mn}^{(j)\pm}(\omega)=\delta_{m,n}\pm \hat{s}_{1/t_{j}}\left(\delta_{m,n-1}+\delta_{m,n+1}\right)
\end{equation}
and following \cite{Babichenko:2003rf}, we multiply the saddle equations
above by the matrix $Q_{m,n}^{(j)-}$ and obtain
\begin{eqnarray}\label{eq:BK-TBAnumerics}
\phi_{m}^{(j)}\left(x\right) & = & \delta^{j,1}\delta_{m,K}\frac{1}{T}\arctan\left(e^{\pi x}\right)
\nonumber \\
&& 
-s*\left[L_{-,m-1}^{(j)}+L_{-,m+1}^{(j)}-L_{+,m}^{(j-1)}-L_{-,m}^{(j+1)}\right]\left(x\right)
\qquad j<K-1\nonumber \\
\phi_{m}^{(K-1)}\left(x\right) & = & -s*\left[L_{-,m-1}^{(K-1)}+L_{-,m+1}^{(K-1)}-L_{+,m}^{(K-2)}\right]\left(x\right)+\left[\frac{s}{s_{1/2}}*Q_{2m,n}^{(K)+}*L_{+,n}^{(K)}\right]\left(x\right)
 \nonumber\\
\phi_{2m}^{(K)}\left(x\right) & = & -s_{1/2}*\left[L_{-,2m-1}^{(K)}+L_{-,2m+1}^{(K)}-L_{+,m}^{(K-1)}\right]\left(x\right)  \\
\phi_{2m-1}^{(K)}\left(x\right) & = & -s_{1/2}*\left[L_{-,2m-2}^{(K)}+L_{-,2m}^{(K)}\right]\left(x\right)\nonumber 
\;.
\end{eqnarray}
We now consider the functions
\begin{equation}
\hat{y}^{(j)}\left(\omega\right)=\sum_{m>0}\hat{a}_{m/t_{j}}\left(\omega\right)\hat{L}_{+,m}^{(j)}(\omega)\label{eq:ydef}
\end{equation}
and we multiply the system (\ref{eq:BK-SaddlePoint}) on the left
by the function $a_{m/t_{j}}$ and sum over $m$, obtaining:
\begin{equation}
\frac{\delta^{j,1}}{T}s*E_{1}^{(1)}-s_{1/t_{j}}*a_{m/t_{j}}*L_{-,m}^{(j)}+\hat{\tilde{M}}^{j,l}*\left(y^{(l)}-s_{1/2}\delta^{j,K-1}\delta^{l,K}L_{+,1}^{(K)}\right)=0
\;,
\end{equation}
with the matrix
\begin{eqnarray}
\hat{\tilde{M}}^{j,l}\left(\omega\right) & = & \frac{\coth\frac{\omega}{2t_{j,l}}}{\coth\frac{\omega}{2t_{j}}}\hat{M}^{j,l}\left(\omega\right)\label{eq:defMhat}
\;.
\end{eqnarray}
The following step is to compute $\tilde{A}=\tilde{M}^{-1}$, the inverse
of the matrix (\ref{eq:defMhat}), and solve for $y^{(j)}$. In particular
\begin{eqnarray}
\hat{\tilde{A}}^{j,l}(\omega) & = & \frac{s\left(\frac{\omega}{2}\right)}{s\left(\omega\right)}\frac{f_{K}(\omega)}{s\left(2\left(K-\max\left(j,l\right)\right)-1\right)}\frac{\sinh\left(\frac{\min\left(j,l\right)\omega}{2}\right)}{\sinh\left(\frac{\omega}{2}\right)}\nonumber \\
\hat{\tilde{A}}^{K,l}(\omega) & = & f_{K}(\omega)\frac{s\left(\frac{\omega}{2}\right)}{s\left(\omega\right)}\frac{\sinh\left(\frac{l\omega}{2}\right)}{\sinh\left(\frac{\omega}{2}\right)}\label{eq:BK-Atilde}\\
\hat{\tilde{A}}^{j,K}(\omega) & = & \frac{f_{K}(\omega)}{s\left(\frac{\omega}{2}\right)}\frac{\sinh\left(\frac{j\omega}{2}\right)}{\sinh\left(\frac{\omega}{2}\right)}\nonumber \\
\hat{\tilde{A}}^{K,K}(\omega) & = & f_{K}(\omega)\frac{\sinh\left(\frac{K\omega}{2}\right)}{\sinh\left(\frac{\omega}{2}\right)}\nonumber 
\;,
\end{eqnarray}
with $f_{K}(\omega)=\frac{\cosh\frac{\omega}{4}}{\cosh\frac{(2K-1)\omega}{4}}$
(see also \cite{Babichenko:2003rf}). Now we substitute our findings
into equation (\ref{eq:BK-freeNrg-long}) and isolate the effective
temperature dependence of the junction free energy: 
\begin{eqnarray}
F_{J}
&=&E_{J}^{(0)}-T\sum_{l=1}^{K}\intop_{\mathbb{R}}d\omega e^{i\omega/\lambda}
\hat{\tilde{A}}^{K,l}\left(\omega\right)\hat{s}_{1/t_{l}}\left(\omega\right)\hat{L}_{-,1}^{(l)}\left(\omega\right)
\nonumber\\
&& \qquad +T\intop d\omega\frac{e^{i\omega/\lambda}\sinh\frac{(K-1)\omega}{2}}{2\cosh\frac{(2K-1)\omega}{4}\sinh\frac{\omega}{2}}
\hat{L}_{+,1}^{(K)}\left(\omega\right)
\label{eq:BK-freeNrg}
\;,
\end{eqnarray}
 with the pseudoenergies $\phi^{(l)}_m$
satisfying the system of equations (\ref{eq:BK-TBAnumerics}).
The temperature-independent part of the junction free energy is:
\begin{equation}\label{eq:BK-GSnrg}
E_{J}^{(0)}=\intop_{\mathbb{R}}d\omega\frac{e^{-|\omega|}\sin\frac{\omega}{\lambda}}{2\omega\cosh\frac{(M-2)\omega}{4}}-\frac{\pi}{2}
\;,
\end{equation}
having substituted the definition (\ref{eq:BK-Atilde}).

\subsection{The ground state energy}\label{sec:GSenergy}

We have summarized the results for the shift of ground state energy due
to the coupling of $M\ge3$ wires with the Majorana bound states
in equation (\ref{eq:GroundStateNRG})
which is of the same form of the impurity contribution in the Kondo
effect (see e.g. \cite{Andrei83}).
This expression can be obtained for even and for odd $M$ by making use 
of a standard integral representation of the logarithm of the $\Gamma$-function
\cite{abramowitz1964handbook} into the integral expressions
(\ref{eq:DK-GSnrg}) and (\ref{eq:BK-GSnrg}).

\begin{figure}
\begin{centering}
\includegraphics[width=0.45\textwidth]{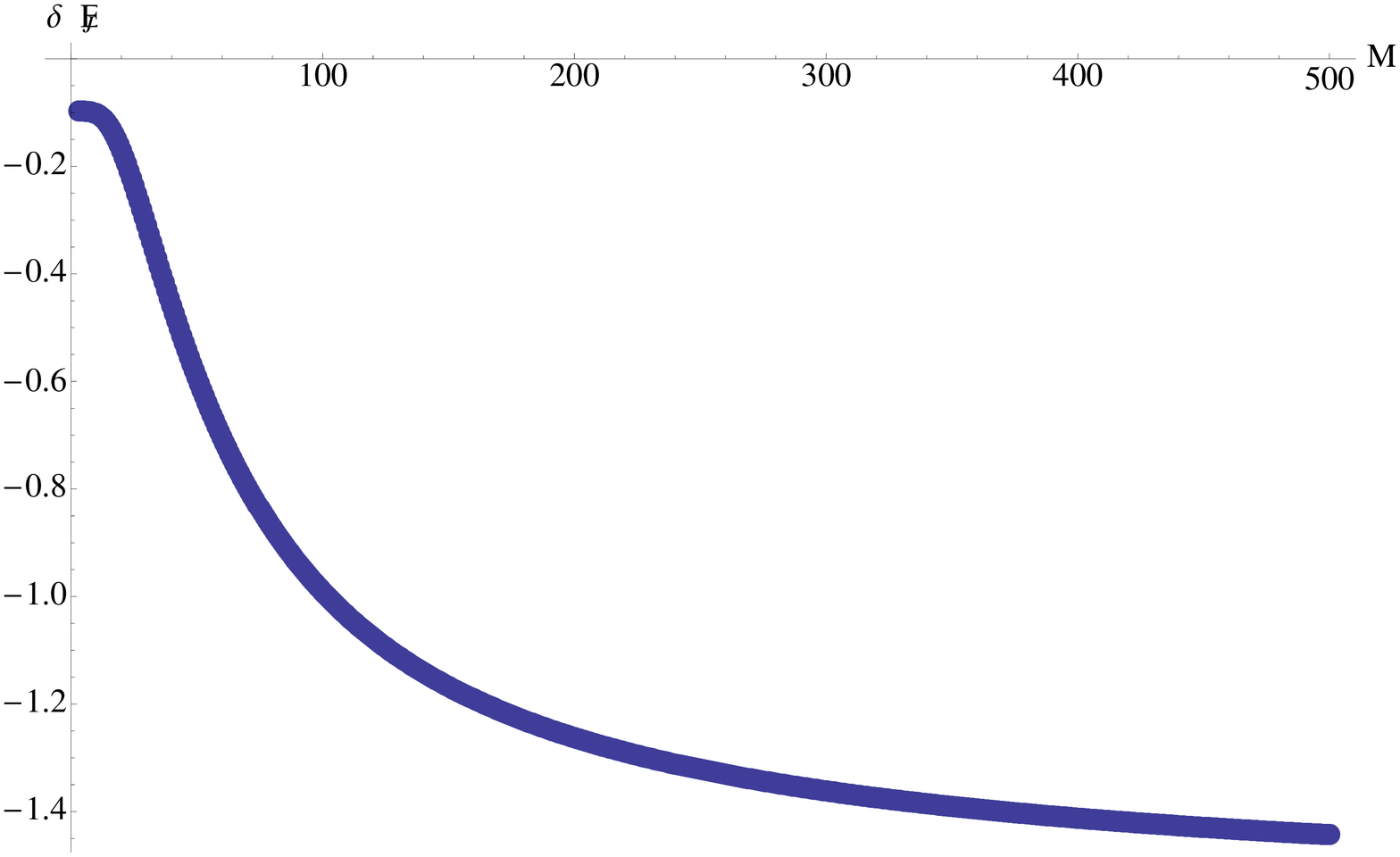}
\includegraphics[width=0.45\textwidth]{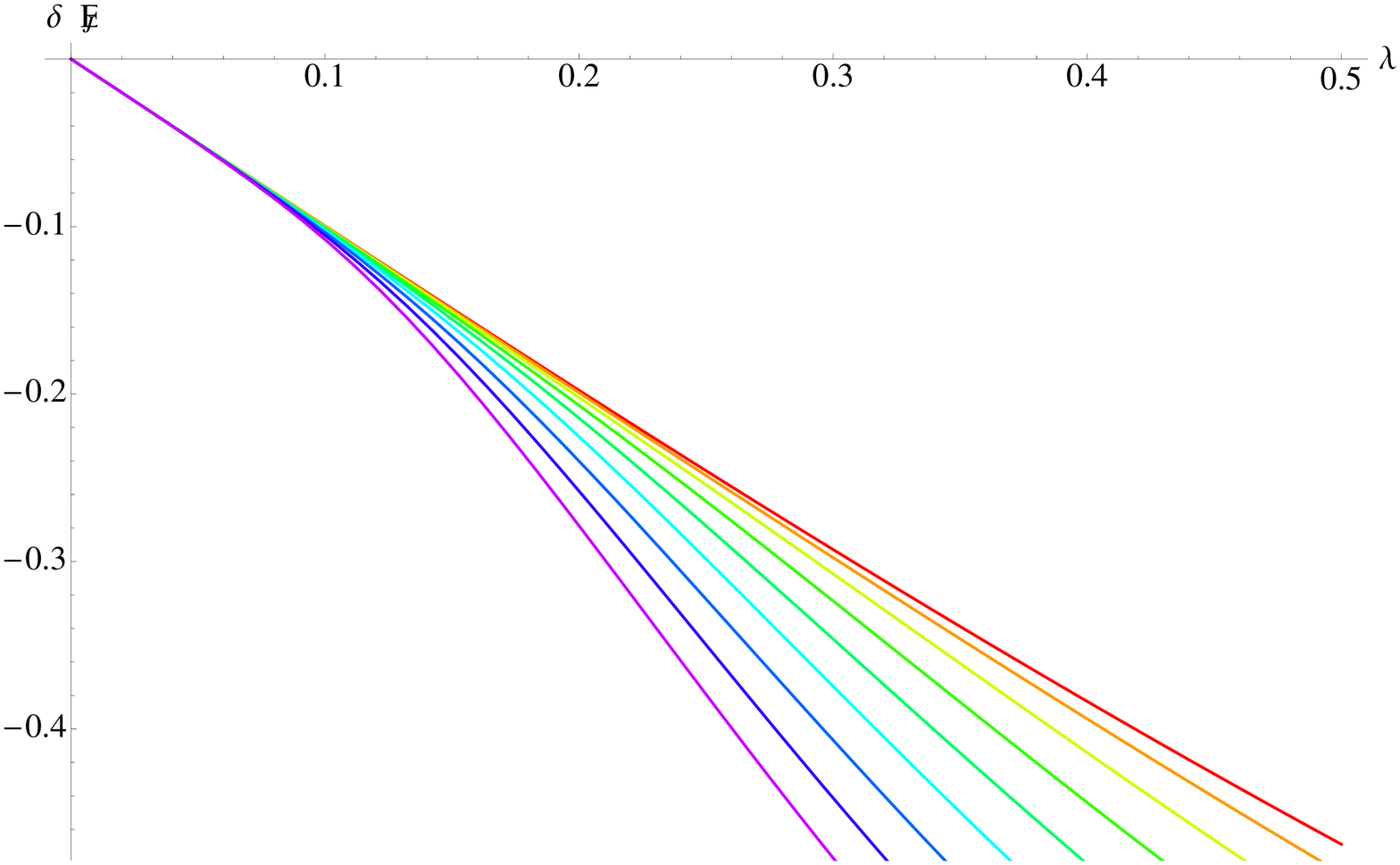}
\par\end{centering}
\caption{Left: ground state energy shift due to the coupling between legs and MEMs, as
a function of the number of legs $M$, for $\lambda=0.1$. Right:
same quantity as a function of $\lambda$ for $M$ between $3$ (top
line, red) and $11$ (bottom line, purple). \label{fig:GSNRG}}
\end{figure}

When the coupling between legs
$\lambda$ goes to zero, the junction energy vanishes linearly in
the coupling with the Majorana box
\begin{equation}
E_{J}^{(0)}(\lambda\to0,M)\sim-\lambda
\end{equation}
with a coefficient which is independent of the number of legs. In
figure \ref{fig:GSNRG}, we show some example of (\ref{eq:GroundStateNRG})
as a function of $M$ and $\lambda$. It is interesting to note that
the next non vanishing order in the expansion is $\lambda^{3}$, with
coefficient $\frac{4+12M-3M^{2}}{48}$ depending on the number of
branches.

\begin{figure}
\begin{centering}
\includegraphics[width=0.45\textwidth]{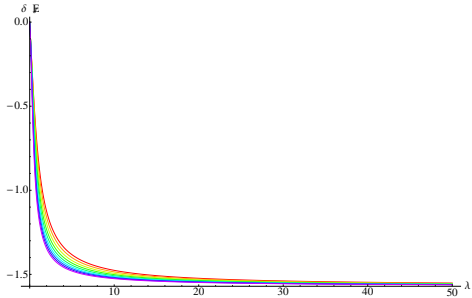}
\par\end{centering}
\caption{Ground state energy shift due to the coupling between the wires and 
the localized Majorana fermions as
a function of $\lambda$ for $M$ between $3$ (top
line, red) and $11$ (bottom line, purple). \label{fig:GSNRGfull}}
\end{figure}

In the strong coupling limit $\lambda\to\infty$, the energy goes to 
 a constant value as
\begin{equation}
E_{J}^{(0)}(\lambda\to\infty,M)  \sim
- \frac{\pi}{2}
+\frac{2}{M-2}\left(\psi\left(\frac{3M-2}{4(M-2)}\right)-\psi\left(\frac{M+2}{4(M-2)}\right)\right)\frac{1}{\lambda}
\;,
\end{equation}
with $\psi$ being the logarithmic derivative of the $\Gamma$ function.
This quantity is monotonous in the coupling and is plotted 
in Figure \ref{fig:GSNRGfull}. Here, it should be noted that 
there exists a point at intermediate $\lambda$ where
 the scaling with $M$ changes.

\section{The entropy of the Majorana edge modes}\label{sec:Entropy}
In this section, we compute the entropy at $T\to\infty$ of the MEMs
and determine the dimension of the Hilbert space $\mathcal{}{H}_J$ of the 
additional Majorana degrees of freedom localized at the 
concurring ends of the wires. We will compute the infinite-temperature limit of the entropy $S_J^{(\infty)}$
of the central region by looking at the asymptotic behaviour of the free energy of the localized Majorana modes,
which in this limit is indeed dominated by the entropy and goes as 
$F_J(T\to\infty)\sim -TS_J^{(\infty)} =-T\log\left(\dim\mathcal{H}_J\right)$.

In addition, we compute the entropy of the ground state for $T=0$. The residual entropy
found at zero temperature shows that
the degrees of freedom introduced by the Majorana modes,
are in a non-trivial superposition of degenerate states.
We will obtain the entropy $S_J^{(0)}$ of the central region at $T=0$ from the first-order
temperature correction to the free energy \cite{Jerez1998}
\begin{equation}\label{FreeNRG-LowTxpansion}
 F_J(T) = E^{(0)}_J - T S_J^{(0)} + O\left(T^{1+2\frac{M-2}{M}}\right)
 \;,
\end{equation}
which in turn is obtained by substituting the zero-temperature solutions to the
TBA systems (\ref{eq:DK-phi-numerics}) and (\ref{eq:BK-TBAnumerics}).

Similarly to what observed in \cite{Babujian1983317,takahashiBook}, the systems 
(\ref{eq:DK-phi-numerics}) and (\ref{eq:BK-TBAnumerics}) admit constant solutions
\cite{KirillovReshetikhin,Kuniba1993Thermodynamics} both when $T\to0$ and when $T\to\infty$.
These solutions satisfy a set of coupled algebraic
equations which we call ``asymptotic'' TBA. These are a simplified version
of the TBA equations: we introduce them in section \ref{sec:EntropyEven} for even $M$
and in section \ref{sec:EntropyOdd} for odd $M$.
Remarkably, the solution of this system can be parametrized in a similar way in the 
low- and in the high-temperature limits. Therefore, we will be able to study the high-temperature limit in
sections \ref{sec:EntropyEvenHigh} and \ref{sec:EntropyOddHigh} for even and for odd 
number of wires, respectively. We will also be able to study the low-temperature limit in 
sections \ref{sec:EntropyEvenLow} and \ref{sec:EntropyOddLow}, respectively for even and for
odd values $M$.
The following sections will be devoted to the analysis of the TBA equations for
even 
and odd
number of branches.

\subsection{The asymptotic TBA: even number of branches}\label{sec:EntropyEven}
Let us analyse the system (\ref{eq:DK-phi-numerics}): due to the boundedness of the source term,
located in the equation for $\phi^{(1)}_2$, the infinite temperature limit removes such 
term from the system. As a consequence, a constant solution $\bar\phi^{(j)}_m$ appears.
The presence of such solution, which only occurs in the limits of zero and at infinite temperature,
greatly simplifies the analysis of the TBA equations.
In fact, we plug such constant ansatz into (\ref{eq:DK-phi-numerics}) and rewrite the system in
terms of the quantities
\begin{equation}\label{eq:eta-definition}
 \bar\eta^{(j)}_m \equiv e^{-\bar\phi^{(j)}_m}
 \;,
\end{equation}
obtaining
\begin{eqnarray}\label{eq:DK-asymptoticTBA}
\left(\bar\eta^{(1)}_{m}\right)^2&=&  \frac{\left(1+\bar\eta^{(1)}_{m-1}\right)\left(1+\bar\eta^{(1)}_{m+1}\right)}
                                        {\left(1+1/\bar\eta^{(2)}_{m}\right)}\nonumber\\
 \left(\bar\eta^{(j)}_{m}\right)^2 &=& \frac{\left(1+\bar\eta^{(j)}_{m-1}\right)\left(1+\bar\eta^{(j)}_{m+1}\right)}
                                        {\left(1+1/\bar\eta^{(j-1)}_{m}\right)\left(1+1/\bar\eta^{(j+1)}_{m}\right)} \nonumber\\  
 \left(\bar\eta^{(K-2)}_m \right)^2&=& \frac{\left(1+\bar\eta^{(K-2)}_{m-1}\right)\left(1+\bar\eta^{(K-2)}_{m+1}\right)}
                                        {\left(1+1/\bar\eta^{(K-3)}_{m}\right)\left(1+1/\bar\eta^{(K-1)}_{m}\right)\left(1+1/\bar\eta^{(K)}_{m}\right)} \nonumber\\
 \left(\bar\eta^{(K-1)}_m \right)^2=\left(\bar\eta^{(K)}_m \right)^2&=&\frac{\left(1+\bar\eta^{(K)}_{m-1}\right)\left(1+\bar\eta^{(K)}_{m+1}\right)}
                                        {\left(1+1/\bar\eta^{(K-2)}_{m}\right)}
                                        \;.
\end{eqnarray}
It is implicit, here, that $\bar\eta^{(j)}_{0}$ for $j=1,\ldots,K$.
The solution of this system corresponds to the asymptotic value of the solution of 
the TBA equations at $x\to-\infty$, when the source term disappears.

On the other hand, when the temperature goes to zero,
the positive-defined driving term diverges to $+\infty$ for all the values of the argument.
As a result, it is possible to substitute another rapidity-independent ansatz into the TBA equations
and one obtains once again the system (\ref{eq:DK-asymptoticTBA}), this time with
the boundary condition
\begin{equation} \label{eq:asyTBA-lowTconstraint}
 \bar\eta^{(1)}_2 = 0 \qquad\qquad (T\to0)
 \;,
\end{equation}
coming from the definition (\ref{eq:eta-definition}). 
On the contrary, at infinite temperature, since there are no divergences that can force one order of solutions
to vanish, we expect $\bar\eta^{(j)}_n\ne0$ for all values of $j$ and $m$.
Therefore, the very same set of algebraic equations (\ref{eq:DK-asymptoticTBA}) holds in the 
two regimes, high and low temperature, yet these regimes are distinguished by the boundary conditions
(\ref{eq:asyTBA-lowTconstraint}) only.

Once we have the asymptotic solutions at infinite or at zero temperature, we insert them
into the formula for the free energy (\ref{eq:DK-JunctionFreeEnergy}).
Then the constant asymptotic solution factors out of the integral, and one obtains
that the temperature dependence of the free energy is given by
\begin{equation}\label{eq:DK-asymptoticFreeEnergy}
 F_J(T)\sim \frac{T}{2}\left[\sum_{l=1}^{K-2} l \log f^{(l)}_1 +\left(\frac{K}{2}+\frac{K-2}{2}\right)\log f^{(K)}_1\right]
 \qquad\qquad (M=2K)
 \;,
\end{equation}
in which there appear the filling fractions
\begin{equation}\label{eq:f-definition}
 f^{(j)}_m=\frac{1}{1+\bar\eta^{(j)}_m}=\frac{1}{1+e^{-\bar\phi^{(j)}_m}}
 \;.
\end{equation}
Note that (\ref{eq:DK-asymptoticFreeEnergy}) represents both the leading
contribution to the free energy as $T\to\infty$ and the first order correction 
in a low-temperature expansion. Of course, the asymptotic solutions for the 
$f^{(j)}_1$s will be different in the two cases.

The solution to (\ref{eq:DK-asymptoticTBA}) is given in \cite{Kuniba1993Thermodynamics,KirillovReshetikhin} by the expression:
\begin{equation}\label{eq:fofQ}
 f^{(j)}_m\equiv\frac{1}{1+\bar\eta^{(j)}_m}=1-\frac{Q^{(j)}_{m-1}Q^{(j)}_{m+1}}{\left(Q^{(j)}_m\right)^2}
 \;.
\end{equation}
The quantities $Q^{(j)}_{m}$ are objects which only depend on the overall symmetry of the problem and
satisfy \cite{Kirillov} the system of equations
\begin{eqnarray}\label{eq:DK-Qsystem}
 \left(Q^{(j)}_m\right)^2-Q^{(j)}_{m-1}Q^{(j)}_{m+1}&=&Q^{(j+1)}_{m}Q^{(j-1)}_{m}\qquad\qquad 0<j<K-2  \nonumber\\
  \left(Q^{(K-2)}_m\right)^2-Q^{(K-2)}_{m-1}Q^{(K-2)}_{m+1}&=&Q^{(K)}_{m}Q^{(K-1)}_{m}Q^{(K-3)}_{m}   \\
 \left(Q^{(K-1)}_m\right)^2-Q^{(K-1)}_{m-1}Q^{(K-1)}_{m+1}&=&Q^{(K-2)}_{m}  \nonumber\\
 Q^{(K-1)}_m &=& Q^{(K)}_m 
\end{eqnarray}
with initial conditions
\begin{eqnarray} \label{eq:DK-Q_1}
 Q^{(j)}_1&=& \sum_{l=0}^{\left\lfloor(j-1)/2\right\rfloor}\chi_{\omega_j-2l}(y\rho)  \qquad\qquad 0<j<K-1 \nonumber\\
Q^{(K-1)}_1=Q^{(K)}_1&=& \chi_{\omega_{K-1}}(y\rho) = \chi_{\omega_K}(y\rho)
\;.
\end{eqnarray}
It is understood that $Q^{(0)}_m=1$ for all $m$, as well as $Q^{(j)}_0=1$ for all $j$.

The expressions above are written in terms of sums of
characters $\chi_\omega$ of the corresponding Lie algebra representations, each built on a given highest weight $\omega$
described in \cite{Kuniba1993Thermodynamics}.
Such characters are specialized to 
the point $y\rho$, where $\rho=\sum_{j=1}^K \omega_j$ is the Weyl vector,
given by the sum of the fundamental weights $\left\lbrace \omega_j \right\rbrace_{j=1,\ldots,K}$
The complex number $y=\frac{2\pi i}{g+l}$ is determined from
the dual Coxeter number of the symmetry algebra ($g=2K-2$ for the $D_K$ algebra) and the positive
integer $l$; the latter is the order at which the vanishing of the solution $\bar\eta^{(1)}_l$ is imposed,
as described below and in \cite{Kuniba1993Thermodynamics}, and for the sake of definiteness
it will be $l=2$ in the low-temperature limit and $l\to\infty$ in the high-temperature limit.

In order to compute the free energy, we solve for the fillings (\ref{eq:f-definition})
using the system (\ref{eq:DK-asymptoticTBA}):
\begin{eqnarray}\label{eq:DK-f-asy}
f^{(j)}_1 &=& \frac{Q^{(j+1)}_{1}Q^{(j-1)}_{1}}{\left(Q^{(j)}_{1}\right)^2} \qquad \qquad 0<j<K-2  \nonumber\\
f^{(K-2)}_1 &=& \frac{Q^{(K-3)}_{1}Q^{(K-1)}_{1}Q^{(K)}_{1}}{\left(Q^{(K-2)}_{1}\right)^2} \\
f^{(K-1)}_1 = f^{K}_1 &=& \frac{Q^{(K-2)}_{1}}{\left(Q^{(K)}_{1}\right)^2}  \nonumber
\;\;.
\end{eqnarray}
Here below, we specialize the solution to the two limiting cases. These are different
only in the value of $l$, the formulas (\ref{eq:DK-Q_1}) and (\ref{eq:DK-Qsystem})
holding for both the constant ansatze.

\subsection{The high temperature limit}\label{sec:EntropyEvenHigh}
Analogously to the results for the spin-$s$ Heisenberg spin chain and the multichannel Kondo model,
we seek a rational solution to (\ref{eq:DK-asymptoticTBA}), which is non-vanishing
for all the $\bar\eta^{(j)}_m$s.
In particular, this is obtained when $y\to 0$ (or $l\to \infty$). 
Then the characters $\chi_{\omega}(y\rho)$ appearing in (\ref{eq:DK-Q_1})
yield, by definition, the dimension of the 
representation of highest weight $\omega$. In particular, we have:
\begin{eqnarray}\label{eq:DK-dimensionRepresentation}
&& \chi_{\omega_j}(0)=\dim(\omega_j) = \binom{2K}{j}  \nonumber\\
&&\chi_{\omega_K}(0)=\chi_{\omega_{K-1}}(0)=\dim(\omega_{K-1})=\dim(\omega_{K}) = 2^{K-1}
\;.
\end{eqnarray}
By inserting such a solution into (\ref{eq:DK-Q_1}) and the latter into (\ref{eq:DK-f-asy})
we obtain the asymptotic fillings, corresponding to the solution at high temperature
of the TBA system. These are explicitly listed in appendix \ref{sec:xplicit}.
By substituting these into (\ref{eq:DK-asymptoticFreeEnergy}) and evaluating
it numerically for several values of $M$, we conclude that
\begin{equation}\label{eq:DK-dimensionImpurityHilbert}
 \dim \mathcal{H}_J(M)=2^{\frac{M}{2}-1}\;.
\end{equation}
Note that, for $M=4$, (\ref{eq:DK-dimensionImpurityHilbert}) yields the impurity entropy of the $2$-channel Kondo model,
while, for $M=6$, the formula reproduces the impurity entropy of the $SU(4)$ Coqblin-Schrieffer model \cite{Jerez1998}
with impurity in the fundamental representation.
These correspondences are possible only if the dimension $\dim \mathcal{H}_J(M)$
is the dimension of $M$ Majorana fermions projected to a definite fermion number parity sector.
Indeed, four MEMs have a Hilbert space of dimension $\dim \mathcal{H}_M(4)=4$,
while the Hilbert space of a localized spin-$1/2$ magnetic impurity, such as the one characterizing the two-channel Kondo model,
only has dimension $\dim \mathcal{H}_J(4)=2$.
The same reasoning applies to the case $M=6$, for which the Hilbert space $\mathcal{H}_M(6)$
has dimension $\dim \mathcal{H}_M(6)=4$, corresponding to the dimension of the Hilbert space
of six MEMs of given fermion number parity.

\subsection{The low temperature limit}\label{sec:EntropyEvenLow}
As observed earlier, at $T \to 0$ we need to solve (\ref{eq:DK-asymptoticTBA})
with the constraint (\ref{eq:asyTBA-lowTconstraint}).
This is written in terms of the specialized characters of the irreducible representations
of the $D_K$ algebra at the point $\frac{i\pi}{K}\rho$, which corresponds to the choice
$l=2$ in (\ref{eq:DK-Q_1}).
These are provided in \cite{Kirillov} as:
\begin{eqnarray}
\label{DK-qcharacters}
 \chi_{\omega_1}\left(\frac{i\pi\rho}{K}\right) &=& \frac{1}{\sin\frac{\pi (K-1)}{2K}} \frac{\sin\frac{\pi (2K-2)}{2K}}{\sin\frac{\pi }{2K}}  \nonumber\\
%  \chi_{\omega_j}\left(\frac{\pi\rho}{K}\right) &=& \frac{\sin\frac{\pi (K-j)}{K}}{\sin\frac{\pi (K-j)}{2K}\sin\frac{j\pi }{2K}} 
% 						  \prod_{l=1}^{j-1}\frac{\sin\frac{(2K-l)\pi}{2K}}{\sin\frac{l\pi}{2K}}		\qquad\quad 1<j<K-1  \nonumber\\
 \chi_{\omega_K}\left(\frac{i\pi\rho}{K}\right) = \chi_{\omega_{K-1}}\left(\frac{i\pi\rho}{K}\right) &=& \prod_{1\le i<j \le K}\frac{\sin\frac{\pi (2K+1-i-j)}{2K}}{\sin\frac{\pi (2K-i-j)}{2K}}
\;.
\end{eqnarray}
At this particular specialization, these three characters are enough \cite{Kuniba1993Thermodynamics} to determine
the full solution recursively, thanks to the constraint (\ref{eq:asyTBA-lowTconstraint}), which implies the property
$Q^{(j)}_2=1$, $Q^{(j)}_3=0$, $j=1<\ldots,K$, established in \cite{Kuniba1993Thermodynamics}.
In fact, in (\ref{eq:DK-f-asy}), we only need the $Q^{(j)}_1$, $j=1<\ldots,K$.
If $K>3$, we can specialize the first of (\ref{eq:DK-Qsystem}) for $j=1$ and,
using $Q^{(1)}_1=\chi_{\omega_1}\left(\frac{i\pi\rho}{K}\right)$, solve for $Q^{(2)}_1$.
Analogously, we obtain recursively the results up to $j=K-2$, which is most rapidly done by solving numerically the recurrence.

Finally, evaluating numerically the free energy (\ref{eq:DK-asymptoticFreeEnergy}) for several values of $M$,
we conclude that the residual ground state entropy introduced by the interaction of the bulk fermions with the Majorana degrees of freedom is
\begin{equation}\label{eq:DK-residualEntropy}
 S^{(0)}_J=\log \sqrt{\frac{M}{2}}
 \;,
\end{equation}
in agreement with the boundary conformal field theory results of \cite{Altland2014}.

\subsection{The asymptotic TBA: odd number of branches}\label{sec:EntropyOdd}
In the same way as for even $M$, the driving term
disappears from the system (\ref{eq:BK-TBAnumerics}) both 
in the high- and in the low-temperature limits.
Therefore, the TBA system is reduced to:
\begin{eqnarray}\label{eq:BK-asymptoticTBA}
 \left(\bar\eta^{(1)}_{m}\right)^2&=&
 \frac{\left(1+\bar\eta^{(1)}_{m-1}\right)\left(1+\bar\eta^{(1)}_{m+1}\right)}{\left(1+{1}/{\bar\eta^{(2)}_{m}}\right)}
 \nonumber\\
 \left(\bar\eta^{(j)}_{m}\right)^2&=&
 \frac{\left(1+\bar\eta^{(j)}_{m-1}\right)\left(1+\bar\eta^{(j)}_{m+1}\right)}
 {\left(1+{1}/{\bar\eta^{(j-1)}_{m}}\right)\left(1+{1}/{\bar\eta^{(j+1)}_{m}}\right)}
 \qquad\qquad 1<j<K-1
 \nonumber\\
 \left(\bar\eta^{(K-1)}_{m}\right)^2&=&
 \frac{ \left(1+\bar\eta^{(K-1)}_{m-1}\right)\left(1+\bar\eta^{(K-1)}_{m+1}\right)}{ \left(1+{1}/{\bar\eta^{(K-2)}_{m}} \right)
 \left(1+{1}/\bar\eta^{(K)}_{2m}\right)^2\left(1+{1}/{\bar\eta^{(K)}_{2m-1}}\right)\left(1+{1}/{\bar\eta^{(K)}_{2m+1}}\right)}
 \nonumber\\
 \left(\bar\eta^{(K)}_{2m}\right)^2&=& 
 \frac{\left( 1+\bar\eta^{(K)}_{2m-1} \right)\left( 1+\bar\eta^{(K)}_{2m+1} \right)}
 {\left(1+{1}/{\bar\eta^{(K-1)}_{m}}\right)}
 \\
 \left(\bar\eta^{(K)}_{2m-1}\right)^2&=&  \left( 1+\bar\eta^{(K)}_{2m-2} \right)\left( 1+\bar\eta^{(K)}_{2m} \right) 
 \nonumber
\end{eqnarray}
and the two regimes are distinguished by the fact that (\ref{eq:asyTBA-lowTconstraint}) has to be
imposed at $T=0$, while we have $\bar\eta^{(j)}_m\ne0$ for all $1\le j \le K$ and all $m\ge1$
at $T\to\infty$. Note that $\bar\eta^{(j)}_{0}$ for $j=1,\ldots,K$.

The solution to (\ref{eq:BK-asymptoticTBA}) is again given by (\ref{eq:fofQ}), but now
the $Q$s are solutions \cite{Kirillov} of:
\begin{eqnarray}\label{eq:BK-Qsystem}
 \left(Q^{(j)}_m\right)^2-Q^{(j)}_{m-1}Q^{(j)}_{m+1}&=&Q^{(j+1)}_{m}Q^{(j-1)}_{m}\qquad\qquad 0<j<K-1  \nonumber\\
  \left(Q^{(K-1)}_m\right)^2-Q^{(K-1)}_{m-1}Q^{(K-1)}_{m+1}&=&Q^{(K)}_{2m}Q^{(K-2)}_{m}   \\
 \left(Q^{(K)}_m\right)^2-Q^{(K)}_{m-1}Q^{(K)}_{m+1}&=&Q^{(K-1)}_{\left[\frac{m}{2}\right]}Q^{(K-1)}_{\left[\frac{m+1}{2}\right]}  \nonumber
 \;,
\end{eqnarray}
with initial conditions given by
\begin{equation}\label{eq:BK-Q1}
 Q^{(j)}_1=\sum_{l=0}^{\left\lfloor (j-1)/2\right\rfloor}\chi_{\omega_{j-2l}}\left(\frac{2\pi i}{g+l}\rho\right) \quad (0<j<K)\quad,
 \quad Q^{(K)}_1=\chi_{\omega_{K}}\left(\frac{2\pi i}{g+l}\rho\right)
 \;,
\end{equation}
in which the dual Coxeter number is $g=2K-1$ for the $B_K$ algebra.

The solution to (\ref{eq:BK-Qsystem}), in turn, is provided in \cite{Kuniba1993Thermodynamics}.
In order to evaluate the free energy, all we will be needing are the expressions:
\begin{eqnarray}\label{eq:BK-f-asy} 
 f^{(1)}_1  &=& \frac{Q^{(2)}_1}{\left(Q^{(1)}_1\right)^2}  \nonumber\\
 f^{(j)}_1  &=& \frac{Q^{(j-1)}_1Q^{(j+1)}_1}{\left(Q^{(j)}_1\right)^2} \qquad\qquad 1<j<K-1  \nonumber\\
 f^{(K-1)}_1  &=& \frac{Q^{(K-2)}_1 Q^{(K)}_2}{\left(Q^{(K-1)}_1\right)^2} =  \frac{Q^{(K-2)}_1 \left(\left(Q^{(K)}_1\right)^2-Q^{(K-1)}_1\right)}{\left(Q^{(K-1)}_1\right)^2}
\\
 f^{(K)}_1  &=& \frac{Q^{(K-1)}_1}{\left(Q^{(K)}_1\right)^2} 
 \;.
 \nonumber
 \end{eqnarray}
in which we note that only the $K$ numbers $Q^{(j)}_1$ appear and they can be read in (\ref{eq:BK-Q1})
once specialized to a particular value of $l$.
By using the definition (\ref{eq:f-definition}) into the free energy (\ref{eq:BK-freeNrg}),
one obtains
\begin{equation} \label{eq:BK-asymptoticFreeEnergy}
 F_J(T)\sim \frac{T}{2}\sum_{l=1}^{K}l\log f^{(l)}_1-T\frac{K-1}{2}\log\left(1-f^{(K)}_1\right)
\qquad \qquad (M=2K+1)
\end{equation}
for the junction free energy. This holds both as the leading order as $T\to\infty$ and as the
first order in a low-$T$ expansion, provided the corresponding asymptotic fillings $f_1^{(j)}$
are inserted.

\subsection{The high temperature limit}\label{sec:EntropyOddHigh}
We select the solution of (\ref{eq:BK-Qsystem}) with $y\to 0$ in the initial values (\ref{eq:BK-Q1}).
Then, the specialized characters become -- by definition -- equal to the dimension of the corresponding
representation. In particular:
\begin{equation}\label{eq:BK-dimensionRepresentation}
 \chi_{\omega_j}(0)=\dim\omega_j = \binom{2K+1}{j} \qquad 0<j<K\qquad\qquad \chi_{\omega_K}(0)=\dim\omega_K = 2^K
 \;.
\end{equation}
Inserting such solutions into (\ref{eq:BK-f-asy}) we obtain the expressions
listed in appendix \ref{sec:xplicit}.
Plugging them into (\ref{eq:BK-asymptoticFreeEnergy})
and evaluating numerically for several values of (odd) $M$, we conclude that
\begin{equation}\label{eq:BK-dimensionImpurityHilbert}
 \dim\mathcal{H}_J=2^{\frac{M-1}{2}}
\end{equation}
for all odd values of $M$ ($\ge3$).

For $M=3$, the dimension of the Hilbert space of four Majorana fermions
with projection on a given fermion number parity sector is
$\dim\mathcal{H}_J(3)=2$, the same as a for a localized spin-$1/2$
impurity such as the one characterizing the four-channel Kondo model.
A related example has been addressed earlier in \cite{LeeWilczek}, who pointed out that,
in a very similar setting, namely a junction of wires with three MEMs interacting among them,
the Hamiltonian can be written in terms of Pauli matrices as well as in terms of operators
satisfying (\ref{eq:PoissonAlgebra}). However, one expects to find inconsistency
between the two operatorially equivalent formulations, in particular in the degeneracy of the spectrum.
In fact, one can build a one-to-one correspondence between the Hilbert spaces only if the fermion number
parity is considered.

\subsection{The low temperature limit}\label{sec:EntropyOddLow}
As $T\to0$, we need to evaluate the solution of (\ref{eq:BK-asymptoticTBA}) with the constraint (\ref{eq:asyTBA-lowTconstraint}).
Hence, we specialize the characters in (\ref{eq:BK-Q1}) to
the point $\frac{2\pi i}{2K+1}$, corresponding to $l=2$ in the initial values (\ref{eq:BK-Q1}).
From the expressions:
\begin{eqnarray}\label{BK-qcharacters}
 \chi_{\omega_1}\left(\frac{\pi}{2K+1}\rho\right) &= & \frac{\sin\frac{(2K-1)\pi}{2K+1}}{\sin\frac{(K-1/2)\pi}{2K+1}\sin\frac{\pi}{2K+1}} \nonumber\\
 \chi_{\omega_K}\left(\frac{\pi}{2K+1}\rho\right) &= & \prod_{l=1}^K  \frac{\sin\frac{(2l-1)\pi}{2K+1}}{\sin\frac{(2l-1)\pi}{2(2K+1)}}
\end{eqnarray}
and rewriting the constraint (\ref{eq:asyTBA-lowTconstraint}) as $Q^{(j)}_3=0$, $Q^{(j)}_2=1$ for $1\le j<K$ and
$Q^{(K)}_5=0$, $Q^{(K)}_4=1$ as in \cite{Kuniba1993Thermodynamics}, we can solve recursively
the system (\ref{eq:BK-Qsystem}). Once again, this is most efficiently done numerically.
Evaluating the free energy of the MEMs by the use of (\ref{eq:BK-f-asy}) into 
(\ref{eq:BK-asymptoticFreeEnergy}) for several values of $M$,
we conclude that the ground state entropy produced by the topological Kondo term is
\begin{equation}\label{eq:BK-residualEntropy}
 S^{(0)}_J=\log \sqrt{M}
 \;,
\end{equation}
which shows that our thermodynamic Bethe ansatz agrees with the results of \cite{Altland2014}, obtained
by boundary conformal field theory.

\section{The specific heat of the Majorana edge modes}\label{sec:SpecificHeat}
A signature of the non Fermi liquid nature of the strongly coupled fixed point
is given by the next term in the expansion (\ref{FreeNRG-LowTxpansion}). In this section,
we shall see that it is indeed of order $O\left(T^{1+2\frac{M-2}{M}}\right)$.
As a consequence, the Majorana contribution to the specific heat 
behaves at low temperatures as
\begin{equation}\label{eq:specificHeat}
 C_J=-T\frac{\partial^2 F_J}{\partial T^2}\sim \left(\frac{T}{T_K}\right)^{\frac{2(M-2)}{M}}
 \;,
\end{equation}
where the (dimensionless) Kondo temperature $T_K$ depends on the coupling between legs as
\begin{equation}\label{eq:KondoTemperature}
 T_K \sim e^{-\frac{\pi}{\lambda(M-2)}}
 \;.
\end{equation}
A non-integer power is a strong and experimentally detectable
signature of the presence of a non Fermi liquid fixed
point. In particular, it is related to the operator content of the
conformal field theory describing the fixed point at strong coupling,
as explained in \cite{Affleck1991641}.

The power we find is indeed the same as what derived from
the conformal dimension of the leading irrelevant operator
in the conformal perturbation theory of \cite{Altland2013} around the strongly
coupled critical point.
Therefore, we confirm such result from our Bethe ansatz computation. 
Also in this approach, it has its origin in the overall symmetry of the problem only,
since what characterizes the Kondo model in the TBA equations, i.e., the functional 
form of the source term, disappears in the derivation of the $Y$-system. The latter
is uniquely specified by the $SO(M)_2$ symmetry of the model, and the periodicity
of the associated TBA equations is all we have used to determine the power characterizing
the temperature dependence of the specific heat as $T\to0$.

\subsection{Even number of branches}
Following \cite{Jerez1998,Andrei83}, we focus on the rapidity regime $x\to -\infty$,
which introduces the first temperature corrections to the asymptotic values of the TBA functions.
In fact, in this regime the source term is very small, being $2e^{\pi x}$,
so that the ratio with the temperature
in the denominator gives a finite result and the rapidity dependence appears in the solution.
For intermediate and large positive values of the rapidity, instead, the solutions of the TBA
equations quickly approach their asymptotic zero-temperature value.
It is convenient, in this regime, to shift the rapidities as
$x = \xi +\frac{1}{\pi}\log T$, with $T\to0$.

Similarly to (\ref{eq:eta-definition}), we can define
\begin{equation}\label{eq:eta-def-lowT}
 \eta^{(j)}_m\left(\xi\right)=e^{-\phi^{(j)}_m\left(\xi+\frac{1}{\pi}\log T\right)}
\end{equation}
and, starting from (\ref{eq:DK-phi-numerics}), we see that the $\eta^{(j)}_m$s satisfy,
for even $M$ and at low temperature, the following TBA system
\begin{eqnarray}\label{eq:DK-TBA-lowT}
\log\eta_{n}^{(j)}\left(\xi\right) & = & -2e^{\pi\xi}\delta^{j,1}\delta_{n,2}+s*\left[L_{-,n-1}^{(j)}+L_{-,n+1}^{(j)}-L_{+,n}^{(j-1)}-L_{+,n}^{(j+1)}\right](\xi)
\nonumber\\
&& \qquad\qquad \qquad\qquad \qquad\qquad \qquad\qquad  \qquad j<K-2
\nonumber \\
\log\eta_{n}^{(K-2)} \left(\xi\right)& = & s*\left[L_{-,n-1}^{(K-2)}+L_{-,n+1}^{(K-2)}-L_{+,n}^{(K-3)}-L_{+,n}^{(K-1)}-L_{+,n}^{(K)}\right]\left(\xi\right)
 \\
\log\eta_{n}^{(j)} \left(\xi\right)& = & s*\left[L_{-,n-1}^{(j)}+L_{-,n+1}^{(j)}-L_{+,n}^{(K-2)}\right]\left(\xi\right)
\quad\qquad j=K-1,K\nonumber 
\end{eqnarray}
in which
\begin{equation}\label{eq:Llogeta-definition}
 L_{\mp,n}^{(j)}\left(\xi\right)=\log\left(1+\left({\eta^{(j)}_m\left(\xi\right)}\right)^{\pm1}\right)
\end{equation}
and the only temperature dependence is the one implicit in the definition (\ref{eq:eta-def-lowT}). The case $M=2K=4$ is:
\begin{eqnarray*}
\log\eta_{n}^{(1)}\left(\xi\right) & = & -2e^{\pi\xi}\delta_{n,2}+s*\left[L_{-,n-1}^{(1)}+L_{-,n+1}^{(1)}-L_{+,n}^{(2)}\right](\xi)\\
\log\eta_{n}^{(2)}\left(\xi\right) & = & s*\left[L_{-,n-1}^{(2)}+L_{-,n+1}^{(2)}-L_{+,n}^{(1)}\right](\xi)\quad.
\end{eqnarray*}

We now look at the free energy (\ref{eq:DK-JunctionFreeEnergy}) and shift the integration variable as above.
In order to express the temperature in terms of the Kondo temperature (\ref{eq:KondoTemperature}), 
we need to rescale the integration variable as $\xi\to\frac{(M-2)\xi}{\pi}$, which brings
its temperature dependence for $T \ll T_K$ into the form ($M=2K$)
\begin{equation}\label{eq:DK-JunctionFreeEnergy-lowT}
F_J(T) \propto -T
 \sum_{l=1}^K \intop_\mathbb{R} \frac{d\xi}{2\pi} \, 
 h^{(l)} \left(\xi-\log \frac{T}{T_K}\right)
 \log\left(1+\eta^{(l)}_1\left(\frac{(M-2)\xi}{\pi}\right)\right)
\;.
\end{equation}
In order to evaluate this expression, being the TBA equations (\ref{eq:DK-TBA-lowT}) symmetric in the exchange $K\leftrightarrow K-1$,
one only needs the functions
\begin{eqnarray}
&&\hat{h}^{(l)} (\omega) = \frac{\sinh\frac{l\omega\pi}{2(M-2)}}{2\cosh\frac{\pi\omega}{4}\sinh\frac{\pi\omega}{2(M-2)}}
\nonumber\\
&&\hat{h}^{(K-1)} (\omega) + \hat{h}^{(K)} (\omega) = \frac{\sinh\frac{\pi\omega}{4}}{2\cosh\frac{\pi\omega}{4}\sinh\frac{\pi\omega}{2(M-2)}}
\;,
\end{eqnarray}
which come directly from the definitions (\ref{eq:s}) and (\ref{eq:DK-Atilde}).
The case $M=2K=4$ is again written as (\ref{eq:DK-JunctionFreeEnergy-lowT}) with
\begin{equation}
 \hat{h}^{(1)}(\omega) = \frac{1}{1+2\cosh\frac{\pi\omega}{2}} \qquad  \qquad
 \hat{h}^{(2)}(\omega) = \frac{2\cosh\frac{\pi\omega}{4}}{1+2\cosh\frac{\pi\omega}{2}}  \;.
\end{equation}
Having already isolated the term in the free energy which is linear
in the temperature, originating from the asymptotic value of the 
functions $\eta^{(j)}_m$s, we now concentrate on the following term, 
which in turn comes from the low-temperature dependence on the rapidity
of these functions.

Being the source term exponential and given the relation
$s*e^{b x} \propto e^{b x}$, a reasonable ansatz
\cite{Desgranges1985} is that, at low enough temperatures, the exponential
behaviour of the source term is retained and we can assume the form
\begin{equation} \label{eq:etas-lowTansatz}
 \eta^{(j)}_m(x)\simeq \bar\eta^{(j)}_m+c^{(j)}_m e^{b x}
 \;,
\end{equation}
with $\bar\eta^{(j)}_m$ being the already discussed asymptotic values. Under the
ansatz (\ref{eq:etas-lowTansatz}), the free energy becomes
\begin{equation}\label{eq:DK-JunctionFreeEnergy-lowT-2}
 F_J(T)\propto -T\log\sqrt{\frac{M}{2}}
- \left(\frac{T}{T_K}\right)^{\frac{b(M-2)}{\pi}}\sum_{l=1}^K \frac{c^{(l)}_1}{1+\bar\eta^{(l)}_1}\intop_\mathbb{R} 
 \frac{d\xi}{2\pi}h^{(l)} \left(\xi \right)e^{ \frac{b(M-2)\xi}{\pi}}
 \;,
\end{equation}
where the term linear in $T$ has been computed in section \ref{sec:EntropyEven}.

In order to fix the coefficient $b$, we note that from
the TBA systems (\ref{eq:DK-TBA-lowT}) a universal $Y$-system for the
analytic continuation in the complex plane of the functions $\eta^{(j)}_m$
can be derived, as explained in \cite{KunibaReview}.
Universality, in this context, means that the source term of the integral
equations disappear, i.e., that only the $SO(M)_2$ structure of the TBA 
is retained in the $Y$-system, while the source, characterizing the Kondo
models, plays no role.
We refer the reader to the original literature \cite{KunibaReview} for
the explicit form of the $SO(M)_2$ $Y$-system
and just state here that its solutions have periodicity:
\begin{equation}
 \eta^{(j)}_m \left(\xi+i(g+2) \right) = \eta^{(j)}_m \left( \xi \right)
 \qquad\qquad g=M-2
 \;.
\end{equation}
which has been obtained in \cite{Kuniba94,inoue2010} for both even and odd $M$.
It follows (see \cite{ZamolodchikovADE}) that they can be expanded
in powers of the variable $e^{\frac{2 \pi x}{g+2}}$.

Substituting this expansion in the free energy (\ref{eq:DK-JunctionFreeEnergy-lowT-2})
and using (\ref{eq:specificHeat}),
we obtain that the specific heat of the MEMs vanishes at $T\to0$ as
the power law $\left(\frac{T}{T_K}\right)^{\frac{2(M-2)}{M}}$.

For $M=4,6$, this value reproduces the results of \cite{Jerez1998} for the
equivalent Coqblin-Schrieffer models.

\subsection{Odd number of branches}
We repeat here the procedure for an odd number of branches. After the change of 
variable $x=\xi+\frac{1}{\pi}\log T$ we obtain, again at low temperatures and
with the definitions (\ref{eq:eta-def-lowT}) and (\ref{eq:Llogeta-definition}),
the TBA system
\begin{eqnarray}\label{eq:BK-TBAnumerics-lowT}
\log\eta_{m}^{(j)}\left(\xi\right) & = & -2\delta^{j,1}\delta_{m,K}e^{\pi \xi}
+s*\left(L_{-,m-1}^{(j)} +L_{-,m+1}^{(j)}-L_{+,m}^{(j-1)}-L_{-,m}^{(j+1)}\right)\left(\xi\right)
\nonumber \\
&& 
\qquad\qquad\qquad\qquad\qquad\qquad\qquad\qquad\qquad j<K-1\nonumber \\
\log\eta_{m}^{(K-1)}\left(\xi\right) & = & s*\left(L_{-,m-1}^{(K-1)}+L_{-,m+1}^{(K-1)}-L_{+,m}^{(K-2)}\right)\left(\xi\right)
\nonumber \\ &&
-\left(\frac{s}{s_{1/2}}*Q_{2m,n}^{(K)+}*L_{+,n}^{(K)}\right)\left(\xi\right) \nonumber
\\
\log\eta_{2m}^{(K)}\left(\xi\right) & = & s_{1/2}*\left(L_{-,2m-1}^{(K)}+L_{-,2m+1}^{(K)}-L_{+,m}^{(K-1)}\right)\left(\xi\right) \\
\log\eta_{2m-1}^{(K)}\left(\xi\right) & = & s_{1/2}*\left(L_{-,2m-2}^{(K)}+L_{-,2m}^{(K)}\right)\left(\xi\right)
\;.
\nonumber 
\end{eqnarray}

The temperature dependence of the free energy for an odd number of legs and $T \ll T_K$ can then be rewritten as
\begin{eqnarray}\label{eq:BK-JunctionFreeEnergy-lowT}
 F_{J}
&\propto&
-T
\sum_{l=1}^{K}\intop_{\mathbb{R}} \frac{d\xi}{2\pi} h^{(l)}\left(\xi-\log\frac{T}{T_K}\right) \log\left(1+\eta^{(l)}_1 \left(\frac{(M-2)\xi}{\pi}\right)\right)
\nonumber\\
&& 
+T
\intop_{\mathbb{R}} \frac{d\xi}{2\pi} p^{(K)} \left(\xi-\log\frac{T}{T_K}\right)  \log\left(1+\frac{1}{\eta^{(K)}_1 \left(\frac{(M-2)\xi}{\pi}\right)}\right)
\;,
\end{eqnarray}
with the functions:
\begin{eqnarray}
 \hat h^{(l)}(\omega) &=& \frac{\sinh\frac{l\pi\omega}{2(M-2)}}{\cosh\frac{\pi\omega}{4}\sinh\frac{\pi\omega}{2(M-2)}}
\nonumber\\
 \hat h^{(K)}(\omega) &=& \frac{\sinh\frac{\pi(M-1)\omega}{4(M-2)}}{\cosh\frac{\pi\omega}{4}\sinh\frac{\pi\omega}{2(M-2)}}
  \\
 \hat p^{(K)}(\omega) &=& \frac{\sinh\frac{\pi(M-3)\omega}{4(M-2)}}{\cosh\frac{\pi\omega}{4}\sinh\frac{\pi\omega}{2(M-2)}}
 \nonumber
\end{eqnarray}
having been obtained from (\ref{eq:BK-Atilde}) and (\ref{eq:s}).

As explained for even $M$, substituting the ansatz (\ref{eq:etas-lowTansatz})
into (\ref{eq:BK-JunctionFreeEnergy-lowT}), the free energy turns into
\begin{eqnarray}
 F_J(T)&\sim& -T\log\sqrt{M}
-\left(\frac{T}{T_K}\right)^{\frac{(M-2)b}{\pi}}
 \Big[\sum_{l=1}^K\frac{c^{(l)}_1}{1+\bar\eta^{(l)}_1}\intop_\mathbb{R}\frac{d\xi}{2\pi}h^{(l)}\left(\xi\right)e^{\frac{b(M-2)\xi}{\pi}} 
 \nonumber\\
 &&
 \qquad\qquad\qquad
 +\frac{c^{(K)}_1}{\bar\eta^{(K)}_1\left(1+\bar\eta^{(K)}_1\right)}\intop_\mathbb{R}\frac{d\xi}{2\pi}p^{(K)}\left(\xi\right)e^{\frac{b(M-2)\xi}{\pi}}\Big]
 \;,
\end{eqnarray}
whose linear term has been determined in section \ref{sec:EntropyOdd}. This expression,
using (\ref{eq:specificHeat}), proves that the specific heat vanishes as a power law
of $T\to0$.
The power itself is identified through the periodicity of the corresponding $Y$-system,
as found in \cite{Kuniba94,inoue2010}:
\begin{equation}
 \eta^{(j)}_m\left(\xi+i(g+2)\right)=\eta^{(j)}_m\left(\xi\right) \qquad\qquad g=M-2
 \;,
\end{equation}
which proves that the power law is just the one stated in (\ref{eq:specificHeat})
and, therefore, that the system at the strong coupling fixed point cannot be modeled by a Fermi liquid.

For $M=3$, this agrees with the known results for the four-channel Kondo model.
We note in passing that the determination of a the specific heat characteristic
power through the periodicity of the associated $Y$-system is, remarkably, more
general than for the model considered.
For instance, the formula 
$C_J\propto \left(\frac{T}{T_K}\right)^\frac{2g}{g+l}$, which can be proved from the 
periodicity properties assessed in \cite{Kuniba94} too,
works in general for the classes of impurity models.
In fact, for the $l$-channel Kondo model, the level of the Bethe ansatz
is $l$, while the dual Coxeter number for $su(2)$ is $g=2$. Therefore
$C_J\propto \left(\frac{T}{T_K}\right)^\frac{4}{2+l}$, which is an elegant
way of deriving an established result \cite{Tsvelick85,Desgranges1985,Affleck1991641}.
In the $l$-species $su(N)$ Coqblin-Schrieffer model, the dual Coxeter number is $g=N$,
hence the specific heat power law at low temperatures
reads $C_J\propto \left(\frac{T}{T_K}\right)^\frac{2N}{N+l}$, as 
obtained in \cite{Jerez1998}.

\section{Concluding remarks}\label{sec:conclusions}
We have used the thermodynamic Bethe ansatz to provide an exhaustive analysis of the 
thermodynamics of the topological Kondo model.
This model describes physical situations where $M$ external quantum wires 
concur in a central region, where the electrons interact with an extended ``spin'',
formed by spatially distant Majorana modes.
The model has $SO(M)$ symmetry.
Our study provided the thermodynamics of this model for arbitrary $M$ and a 
proof that this model represents a natural description of non-Fermi liquid
quantum critical points associated with the $SO(M)_2$ Wess-Zumino-Witten-Novikov
boundary conformal field theory.
Indeed, a key result of our work is the exact computation of the low temperature
behaviour of the specific heat of the central region. We have found that
\begin{equation*}
 C_J\sim \left(\frac{T}{T_K}\right)^\frac{2(M-2)}{M}\;,
\end{equation*}
where $T_K \simeq e^{-\frac{\pi}{\lambda(M-2)}}$ is the Kondo temperature
and argued that the non-integer power originates only from the symmetry of the
strong-coupling fixed point.

We have derived the exact free energy of the topological Kondo model and provided, in a 
closed form, the exact ground state energy shift due to the presence of the Majorana edge modes localized
in the central region as a function of the effective interaction between the fermions in the external 
wires and the localized modes.

We have computed the entropy associated to the Majorana degrees of freedom localized in the 
central region -- the so-called impurity entropy -- both for $T\to0$ and for $T\to\infty$. Our
analysis provided the exact asymptotic expressions of the impurity entropy of the topological
Kondo model for arbitrary $M$.

For $T\to\infty$, the computation of the entropy determines the dimension of the Hilbert space 
associated to the Majorana edge modes localized in the central region and connected to the external wires.
We found that the dimension of the Hilbert space corresponds to the Hilbert space of $2\left\lceil M/2 \right\rceil$
Majorana modes with projection to a definite fermion parity sector.
This shows that it is always possible to preserve, for any given $M$, fermion parity symmetry.
Thus, conservation of the fermion number parity is an essential feature of the
topological Kondo model.

Despite the topological Kondo Hamiltonian having been derived as an effective low-temperature model,
at higher temperature, the topological Kondo effect is expected to survive as long as there 
are no processes breaking the fermion parity, e.g., single-electron tunnelling to or from the island,
which become relevant.

Throughout this paper, we have assumed that the Majorana bound states in the central region were
sufficiently separated in space so as to allow to neglect the direct coupling between pairs of 
Majorana modes.
The inclusion of such terms should be an interesting extension of our work, since
these terms are known to induce - for sufficiently low temperature - the
$M\to M-2$ crossover between different non Fermi liquid phases and could provide a signature
of the presence of Majorana bound states in the central region.

Networks of quantum wires with Majorana edge modes realize the topological Kondo effect
only when the superconducting central region is capacitively coupled with the ground \cite{BeriCooper2012,AltlandEgger,Beri2013,Zazunov14};
when the superconducting central region is floating, the system exhibits remarkable even-odd
effect in the tunnelling conductance through the central region \cite{ZazunovEvenOdd}.
Both situations can be tested experimentally with present day technology, leading to the 
exciting possibility of an experimental test of exact results stemming from Bethe ansatz.

Ising spin chain realizations of the topological Kondo model have been also recently investigated
in \cite{TsvelikIsing}. It was argued that this kind of system exhibits Fermi liquid
behaviour at the strong-coupling fixed point when the number of legs is even, while it produces
a non-Fermi liquid when the number of legs is odd. This feature, due to the reality of the 
fermions in the bulk, is manifest, e. g., in the low temperature behaviour of the specific heat
and marks a difference with the model analysed in this paper, which reaches a
non Fermi liquid fixed point at low temperatures for all values of $M$.

Networks of $XX$ chains could provide a basis for constructing alternative experimental realizations
of the topological Kondo model investigated in this paper: their implementation could be realized 
in ultra-cold atom platforms and we think that it would be a very interesting line of future research
to investigate feasibility and detection schemes in such setups.

\vspace{1cm}
{\it Acknowledgements:}
We would like to thank 
N. Cramp\'e, D. Cassettari, G. Dossena, R. Egger, A. Kuniba, A. Ferraz, 
A. Sedrakyan and A. Tsvelik
for useful discussions and suggestions, as well as Nordita for hospitality
during the program Quantum Engineering of States and Devices.
The work has been carried on with the aid of the high performance computer
system of the International Institute of Physics - UFRN, Natal, Brazil
P.S. and F.B. acknowledge financial support from the Ministry of Science,
Technology and Innovation of Brazil.
P.S. thanks the Ministry
of Science, Technology and Innovation of Brazil MCTI and
UFRN/MEC for financial support and CNPq for granting a
Bolsa de Produtividade em Pesquisa.
V.K. is supported by the grant $DMS-1205422$ and H.B. acknowledges
support from the Armenian grant $11-1c028$ and the Armenian-Russian 
grant $AR-17$.
H.B. and V.K. are grateful to the International
Institute of Physics of UFRN (Natal) for hospitality.

\newpage{}

\begin{appendix}
 \section{Some explicit expressions for the asymptotic filling fractions} \label{sec:xplicit}
 For the sake of clarity, we provide below the expressions for the first-order fillings at high temperature,
 which are connected with the solution of the TBA equations through (\ref{eq:f-definition}), and give some
 more detail on the low-temperature asymptotic fillings as well. These
 directly follow from the procedure explained in section \ref{sec:Entropy}.
 
 \subsection{Even number of branches}
 At high temperature, as illustrated in the main text, we substitute the dimension of the fundamental
 highest-weight representations (\ref{eq:DK-dimensionRepresentation}) into (\ref{eq:DK-Q_1}), thus
 obtaining all the $Q_1^{(j)}$ for $1\le j \le K$. Then we consider the system (\ref{eq:DK-Qsystem}) with $m=1$,
which allows to compute all the $Q_2^{(j)}$ for $1\le j \le K$.
Without any further iteration, we are provided an asymptotic high-temperature solution for the filling
fractions $f_1^{(j)}$ by the use of (\ref{eq:fofQ}).
This solution is (\ref{eq:DK-f-asy}), which reads explicitly:
 \begin{eqnarray}\label{eq:DK-f-high} 
f^{(1)}_1 &=& \frac{\binom{2K}{2}}{\binom{2K}{1}^2} = \frac{(2K)(2K-1)}{2(2K)^2}=\frac{2K-1}{4K}  \nonumber\\
f^{(j)}_1 &=& \frac{\sum_{l=0}^{\left\lfloor (j-2)/2\right\rfloor}\binom{2K}{j-1-2l}\sum_{l=0}^{\left\lfloor j/2\right\rfloor}\binom{2K}{j+1-2l}}
{\left(\sum_{l=0}^{\left\lfloor (j-1)/2\right\rfloor}\binom{2K}{j-2l}\right)^2}
\qquad \qquad 1<j<K-2 \nonumber\\ 
f^{(K-2)}_1 &=& 2^{2(K-1)}\frac{\sum_{l=0}^{\left\lfloor (K-4)/2\right\rfloor}\binom{2K}{K-3-2l}}{\left(\sum_{l=0}^{\left\lfloor (K-3)/2\right\rfloor}\binom{2K}{K-2-2l}\right)^2}
\\
f^{(K-1)}_1 = f^{(K)}_1  &=& 2^{-2(K-1)}\binom{2K}{K-2}
\;.
\nonumber
\end{eqnarray}
Note that these values provide the occupation at high temperature of the real solutions of the Bethe equations by the use of (\ref{eq:PhiDef}).

The low-temperature limit is greatly simplified by the condition 
$Q^{(j)}_2=1$, $Q^{(j)}_3=0$, $j=1<\ldots,K$, discussed in section \ref{sec:EntropyEvenLow}.
In fact, we know $Q^{(1)}_1$ from (\ref{eq:DK-Q_1}) and (\ref{DK-qcharacters}) and we can use it as an initial condition
in order to start a numerical recursive solution of the system (\ref{eq:DK-Qsystem}) with $m=1$.
In this way, we derive
\begin{eqnarray}
 Q^{(2)}_1&=&\left[\chi_{\omega_1}\left(\frac{i\pi\rho}{K}\right)\right]^2-1  \nonumber\\
 Q^{(3)}_1&=&\frac{\left[Q^{(2)}_1\right]^2-1}{Q^{(2)}_1}  \qquad \ldots
\end{eqnarray}
and so on for $Q_1^{(4)}$, ... , until having determined $Q^{(K-2)}_1$. We perform this recursion numerically.
We already know $Q^{(K-1)}_1=Q^{(K-1)}_1$ from (\ref{eq:DK-Q_1}) and (\ref{DK-qcharacters}).
By plugging the solution of this recurrence into (\ref{eq:fofQ}), we obtain the asymptotic filling fractions.

 \subsection{Odd number of branches}
In the high-temperature limit, we have all the $Q_1^{(j)}$ for $1\le j \le K$ by
substituting the dimension of the fundamental highest-weight representations (\ref{eq:BK-dimensionRepresentation})
into (\ref{eq:BK-Q1}). We now consider the system (\ref{eq:BK-Qsystem}) with $m=1$. This
allows to compute all the $Q_2^{(j)}$ for $1\le j \le K$.
Without any further effort, we can already compute all the $f_1^{(j)}$ for $1\le j \le K$, by means of (\ref{eq:fofQ}):
 \begin{eqnarray}\label{eq:BK-f-high} 
 f^{(1)}_1  &=& 
 \frac{\binom{2K+1}{2}}{\binom{2K+1}{1}^2} =\frac{2K}{2(2K+1)} \nonumber\\
 f^{(j)}_1  &=&
 \frac{\sum_{l=0}^{\left\lfloor(j-2)/2\right\rfloor}\binom{2K+1}{j-1-2l} \sum_{l=0}^{\left\lfloor j/2\right\rfloor}\binom{2K+1}{j+1-2l}} 
  {\left(\sum_{l=0}^{\left\lfloor (j-1)/2\right\rfloor}\binom{2K+1}{j-2l}\right)^2}\qquad\qquad 1<j<K-1  \nonumber\\
 f^{(K-1)}_1  &=& 
\frac{2^{2K}-\sum_{l=0}^{\left\lfloor(K-2)/2\right\rfloor}\binom{2K+1}{K-1-2l}}{\left(\sum_{l=0}^{\left\lfloor(K-2)/2\right\rfloor}\binom{2K+1}{K-1-2l}\right)^2}
\sum_{l=0}^{\left\lfloor(K-3)/2\right\rfloor}\binom{2K+1}{K-2-2l}
\\
 f^{(K)}_1  &=& 
 2^{-2K}\sum_{l=0}^{\left\lfloor(K-2)/2\right\rfloor}\binom{2K+1}{K-1-2l}
 \;.
 \nonumber
 \end{eqnarray}
 
In the low-temperature asymptotic limit, we know $Q^{(1)}_1$ by means of (\ref{eq:BK-Q1}) and (\ref{BK-qcharacters}).
Moreover, we know that  $Q^{(j)}_2=1$, $Q^{(j)}_3=0$, $j=1<\ldots,K$, as 
discussed in \cite{Kuniba94} and section \ref{sec:EntropyEvenLow}.
We can then use the system (\ref{eq:BK-Qsystem}) with $m=1$ to set up a recurrence and determine numerically
$Q^{(2)}_1 , \ldots ,  Q^{(K-1)}_1$.
The recurrence goes exactly as in the even-$M$ case and starts as:
\begin{eqnarray}
 Q^{(2)}_1&=&\left[\chi_{\omega_1}\left(\frac{2i\pi\rho}{2K+1}\right)\right]^2-1  \nonumber\\
 Q^{(3)}_1&=&\frac{\left[Q^{(2)}_1\right]^2-1}{Q^{(2)}_1}  \qquad \ldots
\end{eqnarray}
and so on. We already know $Q^{(K)}_1$, again from (\ref{eq:BK-Q1}) and (\ref{BK-qcharacters}).
We substitute the solution of the procedure into (\ref{eq:fofQ}) and obtain the desired
asymptotic filling fractions $f^{(1)}_1 , \ldots ,  f^{(K)}_1$.

\end{appendix}

\newpage{}

\bibliographystyle{unsrt}
\bibliography{Y}

\end{document}